\documentclass[aps,prc,showpacs,superscriptaddress,floatfix,amsmath]{revtex4-1}
\usepackage[latin1]{inputenc}
\usepackage[english]{babel}
\usepackage{graphicx,psfrag}
\usepackage{amsmath}
\usepackage{amssymb}
\usepackage{amscd}
\usepackage{eucal}
\usepackage{color}
\usepackage{bm}

\usepackage[bookmarks,colorlinks=true,urlcolor=blue,linkcolor=black,citecolor=blue]{hyperref}

\newcommand{\nmax}{N_{\rm max}}
\newcommand{\hb}{\hbar\Omega}
\newcommand{\Eone}{E_{0,\kappa=0}^{(1)}}
\newcommand{\Etwo}{E_{0,\kappa=0}^{(1+2)}}

\begin{document}
\title{Extrapolation uncertainties in the importance-truncated No-Core Shell Model}
\author{M.K.G. Kruse}
\email{kruse9@llnl.gov}
\affiliation{Physics Department,
   University of Arizona, 1118 E. 4$^{\rm th}$ Street, Tucson, AZ 85721, USA}
   \affiliation{Lawrence Livermore National Laboratory, P.O. Box 808, L-414, Livermore, California 94551, USA}

\author{E.D. Jurgenson}
\affiliation{Lawrence Livermore National Laboratory, P.O. Box 808, L-414, Livermore, California 94551, USA}

\author{P. Navr\'atil}
\affiliation{TRIUMF, 4004 Wesbrook Mall, Vancouver, Canada}
\affiliation{Lawrence Livermore National Laboratory, P.O. Box 808, L-414, Livermore, California 94551, USA}

\author{B.R. Barrett}
\affiliation{Physics Department,
   University of Arizona, 1118 E. 4$^{\rm th}$ Street, Tucson, AZ 85721, USA}

\author{W.E. Ormand}
\affiliation{Lawrence Livermore National Laboratory, P.O. Box 808, L-414, Livermore, California 94551, USA}

\begin{abstract}
\begin{description}

\item[Background]
The Importance Truncated No-Core Shell model (IT-NCSM) has recently been shown to extend theoretical nuclear structure calculations of p-shell nuclei to larger model ($N_{\rm max}$) spaces. The importance truncation procedure selects only relatively few of the many basis states present in a {\em 'large'} $N_{\rm max}$ basis space, thus making the calculation tractable and reasonably quick to perform. Initial results indicate that the procedure agrees well with the NCSM in which a complete basis is constructed for a given $N_{\rm max}$. 

\item[Purpose] 
An analysis of uncertainties in IT-NCSM such as those generated from the extrapolations to the complete $\nmax$ space have not been fully discussed. We present a method for estimating the uncertainty when extrapolating to the complete $\nmax$ space and demonstrate the method by comparing extrapolated IT-NCSM to full NCSM calculations up to $\nmax=14$. Furthermore, we study the result of extrapolating IT-NCSM ground-state energies to $\nmax=\infty$ and compare the results to similarly extrapolated NCSM calculations. A procedure is formulated to assign uncertainties for $\nmax=\infty$ extrapolations.

\item[Method]
We report on $^6 {\rm Li}$ calculations performed with the IT-NCSM and compare them to full NCSM calculations. We employ the Entem and Machleidt chiral two-body N3LO interaction (regulated at 500~MeV/c), which has been modified to a phase-shift equivalent potential by the similarity renormalization group (SRG) procedure. We investigate the dependence of the procedure on the technique employed to extrapolate to the complete $\nmax$ space, the harmonic oscillator energy ($\hbar \Omega$),  and investigate the dependence on the momentum-decoupling scale ($\lambda$) used in the SRG. We also investigate the use of one or several reference states from which the truncated basis is constructed.

\item[Results]
We find that the uncertainties generated from various extrapolating functions used to extrapolate to the complete $\nmax$ space increase as $\nmax$ increases. The extrapolation uncertainties range from a few keV for the smallest $\nmax$ spaces to about 50 keV for the largest $\nmax$ spaces. We note that the difference between extrapolated IT-NCSM and NCSM ground-state energies, however, can be as large as a 100-250~keV depending on the chosen harmonic oscillator energy ($\hbar\Omega$). IT-NCSM performs equally well for various SRG momentum-decoupling scales, $\lambda=2.02$~fm$^{-1}$ and $\lambda=1.50$~fm$^{-1}$. 

\item[Conclusions] 
In the case of $^6$Li, when using the softened chiral NN N3LO interaction, we have determined the difference between extrapolated $\nmax=\infty$ IT-NCSM and full NCSM calculations to be about 100-300~keV. As $\hb$ increases, we find that the agreement with NCSM deteriorates, indicating that the procedure used to choose the basis states in IT-NCSM depends on $\hb$. We also find that using multiple reference states leads to a better ground-state description than using only a single reference state. 
\end{description}
\end{abstract}

\pacs{21.60.De,03.65.Aa}

\maketitle{}

\section{Introduction}
Various techniques have been developed to calculate the binding energies of the lightest nuclei, for which $A \leq 4$. These are the techniques of Faddeev \cite{Faddeev60,Faddeev61}, the Faddeev-Yakubovsky extension \cite{Yak67,Glockle93,Cies}, the correlated Hyper-Spherical Harmonics \cite{Viviani,Barnea}, or the No-Core Shell Model implemented with Jacobi coordinates \cite{Nav00Jac}. These methods all reproduce the experimental binding energies for the Triton and the $\alpha$-particle, provided realistic nuclear interactions are used such as those that contain a three-body force. However, their extension to heavier nuclei is difficult and not well suited for such applications. When one performs nuclear-structure calculations for heavier systems, such as those in the p-shell, the methods of the No-Core Shell Model (NCSM) \cite{Nav98,Nav00,NavRev09}, the Green's Function Monte-Carlo technique (GFMC) \cite{GFMC97,GFMC00,GFMC02,GFMCrev} or the Coupled-Cluster technique \cite{Bartlett07,Hagen08} are easier to implement and computationally more efficient.

All of these techniques have started to place the realm of nuclear structure on a firm {\em ab-initio} footing. With the increases in computational resources coupled with the many-body techniques at our disposal, we can solve the nuclear Hamiltonian through various methods, and are able to reproduce experimental spectra reasonably well \cite{Nav07}. Recently, the community has been able to predict the spectrum of $^{14}$F \cite{Maris10-F14}, which was later verified by experimental results \cite{Goldberg10}. We are also forming a better understanding of the role nuclear interactions in a many-body setting from our many-body calculations. For instance, using the chiral interaction (\cite{EM500}, \cite{Epelbaum02}), it has been clearly demonstrated that the three-nucleon forces appearing at next-to-next leading order (N2LO) are essential for nuclear saturation and are required to reproduce the observed experimental level ordering \cite{Nav07}.

The large model spaces that are typically encountered in nuclear-structure calculations are in part due to the short-range repulsive behavior of the nuclear interaction. Recently, two different unitary methods have been used to 'soften' the short-range repulsive nature of the nuclear interactions. These are the $v_{\rm low-k}$ (see Refs. \cite{Bogner03a,Bogner03b,Bogner07}) and Similarity Renormalization Group (SRG) methods (see Refs. \cite{Jurg09,Jurg11,Bogner10}). In the case of the SRG, a series of unitary transformations are performed on the bare Hamiltonian, dictated by an RG flow equation. A repeated application of the flow equation leads to a decoupling of high- and low-momentum components. The decoupling is characterized by a momentum scale, $\lambda$, which controls the degree of decoupling that the SRG procedure imposes upon the bare interaction. Typical soft interactions in use today have a range of $1.5 < \lambda < 2.5$~fm$^{-1}$. The use of soft interactions in the NCSM changes the {\em rate} of convergence, {\em i.e.}, smaller model spaces are required for convergence than those required for the bare interaction. Furthermore, the SRG and bare interaction NCSM calculations will converge to the same ground-state energy, provided the SRG procedure is carried out in a unitary manner (a discussion on SRG unitarity for the p-shell nuclei can be found in \cite{Jurg11}). In the case of the NCSM, the rate of convergence is also hampered by the use of harmonic-oscillator (HO) single-particle states, since they have the incorrect asymptotic behavior for nuclear bound-state wavefunctions (but offer the best means of separating the intrinsic and center-of-mass degrees of freedom).

To extend NCSM calculations to the mid p-shell region, or perhaps to the start of the sd-shell, is a challenging task, especially if we want fully converged results. By this, we mean calculations that are free of any NCSM parameters, such as a dependence on the size of the many-body basis. In the case of the NCSM, the main difficulty encountered comes from the relatively quick rise in the number of many-body basis states present in the model spaces. Recently, there have been some attempts to overcome this problem by returning to a traditional shell model approach \cite{Lisetsky08}, by projecting the Hamiltonian onto a smaller sub-model space, or by selecting only a small subset of the large number of many-body basis states present \cite{Roth07,Roth09}. The latter method is called the Importance Truncated No-Core Shell Model (IT-NCSM) and will be the topic of this investigation.

The NCSM is characterized by two parameters, $\nmax$ and $\hb$. $N_{\rm max}$ describes the number of oscillator energy quanta available to the $A$ nucleons above the unperturbed ground-state configuration, which in turn defines the number of single-particle states that must be included to form the many-body basis. We will often refer to the model space and $\nmax$ interchangeably. Since we work in a HO single-particle basis, our calculations are also dependent on the chosen $\hbar \Omega$. Both of these parameters will be discussed in detail in Sec. \ref{sec_ncsm}.

The Importance Truncated No-Core Shell Model (IT-NCSM) has recently been used to extend calculations of p-shell nuclei to larger $N_{\rm max}$ spaces, in which it seems that the calculations have converged \cite{Roth11}. In the importance truncation scheme, a small set of basis states is chosen from the full $N_{\rm max}$ space, using a procedure based on multi-configurational perturbation theory. The required $\nmax$ space to reach convergence in the p-shell, in which the complete basis is constructed, may have many billions of basis states. Although there are computers, and some codes, that can handle such a calculation, it is generally still an unfeasible calculation to perform routinely. When the importance truncation procedure is used, the actual number of basis states kept for the same $N_{\rm max}$ space is usually 10-15 million states, a relatively easy calculation to perform. However, the uncertainties that arise from this truncated basis have not been explored in depth. In this paper, we will discuss IT-NCSM calculations of $^6 {\rm Li}$, for which we can compare our IT results to the complete model space up to $\nmax=14$. We begin with the bare NN N3LO interaction \cite{EM500}, which we transform to a phase-shift equivalent form that has been evolved to momentum scales of $\lambda=2.02$~fm$^{-1}$ and $\lambda=1.50$~fm$^{-1}$. For each interaction, we vary the chosen HO energy, $\hbar\Omega$, employing the values $\hbar\Omega=12,16,20,24$~MeV. The IT results are extrapolated in such a way as to (hopefully) recover the complete $\nmax$ space, in which all basis states are present. We investigate the various extrapolation procedures in great depth and discuss the uncertainties on the various extrapolations (see Sec. \ref{sec_extrapolations}). Finally, we also investigate the sensitivity of IT-NCSM to using one (e.g., the ground state) or several reference states (e.g., excited states) in constructing the truncated basis. We will postpone the issues that have been raised concerning the size-extensivity properties of the importance truncation procedure (see Refs. \cite{Dean08,Roth08-reply}) and choose to discuss estimates on the uncertainties of the current implementations of IT-NCSM.

The NCSM will be discussed in detail in Sec. \ref{sec_ncsm}. In Sec. \ref{sec_it}, we will discuss the importance-truncation selection procedure and in the ensuing sections, we will discuss various implications of the method such as the extrapolations used to obtain the ground state energy (see Sec. \ref{sec_extrapolations}), concentrating on making some reasonable estimates of the uncertainties induced in the IT-NCSM calculations. In section \ref{sec_observations}, we present our detailed investigation on the behavior of IT-NCSM calculations, as various parameters are varied. In Sec. \ref{sec_conclusions}, we present our conclusions.

\section{The No-Core Shell Model}
\label{sec_ncsm}

The NCSM is an {\em ab initio} nuclear-structure technique employing realistic two- and three-body potentials. It is similar to the configuration interaction approach, in which a single-particle basis of HO states are used to construct the antisymmetric many-body basis. Two key differences arise though. The NCSM uses realistic interactions, which provides us with a connection to QCD (when EFT potentials are used), whereas the traditional shell-model uses effective interactions, which are typically tuned empirically to experimental data as is done in Ref. \cite{Wildenthal19845}. Secondly, we allow all $A$ nucleons to be active in our model space, in contrast to the traditional shell model, in which only the valence nucleons are active and the core nucleons remain energetically frozen.

Our starting Hamiltonian, which is translationally invariant, is, 

\begin{equation}
\label{ncsm_ham_eq}
H_A = \frac{1}{A}\sum_{i<j}^A \frac{(\vec{p_i}-\vec{p_j})^2}{2m} + \sum_{i<j}^A V_{{\rm NN},ij} + \sum_{i<j<k}^A V_{{\rm NNN},ijk}+\ldots,
\end{equation}
where $m$ is the nucleon mass and $A$ is the number of nucleons present. The first term represents the relative kinetic energy, the second term represents the nucleon-nucleon (NN) interaction, and the third (fourth) term represents the three nucleon (NNN) (and higher-body) interaction. In this work, we are investigating the convergence properties of the IT procedure and not those of the Hamiltonian, thus the details of the NNN (or higher-body) interaction are not required. Furthermore, employing a NNN force would make the calculations significantly more challenging, and would prevent a comparison between IT results and the full NCSM calculations in the largest $\nmax$ spaces. 

Before the advent of soft-interactions, such as those generated by the SRG or $v_{\rm low-k}$, it was necessary to generate an effective interaction by using the Lee-Suzuki method \cite{Okubo54,Suzuki80} for a given $\nmax$ space and a specific value of $\hbar\Omega$ (in the future we will use the notation ($\nmax$,$\hb$) as a short-hand). The effective interaction significantly improves the convergence properties of the NCSM calculations. The softened phase-shift equivalent potentials can be used directly in their bare form, without the need for generating a Lee-Suzuki-type effective interaction (one should keep in mind that both SRG and $v_{\rm low-k}$ interactions are still effective). In other words, we solve the Schroedinger equation given by Eq. (\ref{ncsm_ham_eq}) directly, obtaining the eigenvalues and eigenvectors. Moreover, the two-body matrix elements for any of these soft interactions are identical in various $\nmax$ spaces when $\hb$ is fixed. This allows us to generate one interaction, the $\nmax=14$ interaction at a specified $\hbar\Omega$ value, and use it for all smaller $\nmax$ calculations as well. 

Once an interaction is chosen, the NCSM calculations depend only on $\nmax$ and $\hb$. Since we construct our many-body basis from single-particle HO states, our calculations are dependent on $\hbar\Omega$. Typically, some initial calculations are performed, in which a range of $\hbar\Omega$ values are employed. The resulting ground-state energies depend on the chosen $\hbar\Omega$. However, in large enough $\nmax$ spaces this dependence is often weak for a range of $\hbar\Omega$ values, which leads to a particular value of $\hbar\Omega$ being chosen for the rest of the calculations. In our calculations, we will use a variety of $\hbar\Omega$ values to study the behavior of importance-truncation.

The $A$ nucleons are active in a model space, denoted as $\nmax$, which refers to the number of oscillator quanta the $A$ nucleons may share amongst themselves, above the lowest non-interacting energy configuration allowed by the Pauli-exclusion principle. For example, in $^6 {\rm Li}$, an $\nmax=4$ model space would allow one-nucleon to occupy the $N=5$ HO shell, the other valence nucleon would stay in the $N=1$ shell, and the remaining 4 nucleons would occupy the $N=0$ shell. Alternatively, the two valence nucleons could occupy the $N=3$ shell, whereas the remaining 4 nucleons remain in the $N=0$ shell. 

The most significant advantage of the NCSM basis is that we can exactly separate the intrinsic states from the spurious center-of-mass (CM) states. The factorization of the center-of-mass and intrinsic states is possible if one uses a complete NCSM HO basis as well as a translationally invariant Hamiltonian. In order to distinguish the intrinsic states from the center-of-mass states, one adds to the translationally invariant Hamiltonian (Eq. \ref{ncsm_ham_eq}) a HO center-of-mass term, $\beta H_{CM}$. $H_{CM}= \frac{A \vec{P}^2}{2m} + \frac{1}{2} m A \Omega^2 \vec{R}^2$, in which $\vec{R}=\frac{1}{A}\sum_i^A \vec{r}_i$, is the CM coordinate. This is commonly referred to as the Lawson projection term \cite{Gloeckner74}. The Lawson term shifts the uninteresting CM states on the order of $\beta\hb$ up in the energy spectrum,  leaving the intrinsic states in the low-lying part of the spectrum. In all of our calculations we set $\beta=5$. To reiterate, the separation of CM from the intrinsic states is only guaranteed when the complete NCSM HO basis is employed. In that case, there exists an orthogonal transformation between single-particle and relative-CM co-ordinates \cite{Moshinsky59}. Once the basis is truncated, as is done in IT-NCSM, we no longer have a complete factorization of the intrinsic wavefunction from the CM wavefunction. Calculations by Roth (see Ref.\cite{Roth09b}) have suggested that the CM contamination remains small, on the order of 100~keV. Roth has also shown that once the extrapolation is performed to the complete $\nmax$ space, that $\langle H_{\rm CM} \rangle$ tends to zero, as it intuitively should.

\section{The Importance Truncation Procedure}
\label{sec_it}

The idea of importance truncation in nuclear structure has been adopted from its use in quantum chemistry, in which it has been used with quite some success \cite{Sherrill99}. At the core of the procedure lies a parameter, $\kappa$, that is directly related to the number of many-body basis states kept in a certain $\nmax$ model space. We will briefly present the main results needed and limit technical discussions of the procedure as much as possible. The paper by Roth \cite{Roth09}, as well as the Ph.D dissertation \cite{Kruse-thesis}, present all the technical details of the importance-truncation procedure in nuclear structure, and we refer interested readers to it, if they desire more details than are presented here.

The importance-truncation procedure is based formally on multi-configurational perturbation theory \cite{Surjan04,Rolik03}. In what follows, we will briefly describe the key equations that are used and how they follow from perturbation theory.

\subsection{The Selection of Many-Body Basis States via the Importance Measure, $\kappa_{\nu}$}

Suppose that one wants to perform an $\nmax=12$ calculation for a given nucleus, but, due to limitations set by current computer architectures (or resources), the full-space calculation, in which all the basis states in the $\nmax=12$ space are kept, is not possible. However, let us assume that an $\nmax=10$ calculation is possible, in which one is able to calculate the ground-state wavefunction, which we will denote as $|\Psi_{\rm ref,\nmax=10}\rangle$. As a first-order approximation to the $\nmax=12$ wavefunction, we can estimate the amplitudes of the $\nmax=12$ basis states using first-order perturbation theory.

\begin{equation}
\label{it_eq_wfpt}
 | \psi_{\nmax=12,{\rm IT}}^{(1)} \rangle =  |\Psi_{{\rm ref},\nmax=10}\rangle + \sum_{\nu \in \nmax=12} \frac{\langle \phi_{\nu} | W | \Psi_{{\rm ref},\nmax=10} \rangle} {\epsilon_{\nu}-\epsilon_{\rm ref,sp}}  | \phi_{\nu}  \rangle
\end{equation}

In Eq. (\ref{it_eq_wfpt}), we have explicitly denoted $|\psi_{\nmax=12,{\rm IT}}^{(1)}\rangle$, as the approximate wavefunction of the full space $\nmax=12$ wavefunction. The $|\phi_{\nu}\rangle$ are the $\nmax=12$ many-body basis states. $|\Psi_{{\rm ref},\nmax=10}\rangle$ is our previously calculated reference state, which in our example we assume is the ground-state wavefunction of the $\nmax=10$ space. The two terms in the denominator refer to the single-particle energy level of the corresponding label. In our implementation, we always take $\epsilon_{\rm ref,sp}$ to be the lowest unperturbed energy configuration of the nucleus. In $^6$Li, this corresponds to taking $\epsilon_{\rm ref,sp}=2*\hbar\Omega$, since two valence nucleons occupy the $N=1$ shell. We neglect the zero-point motion of the HO, since we only require the difference in energy of the single-particle states. Furthermore, $\epsilon_\nu=(12+2)\hbar\Omega$ for the basis states in $\nmax=12$. This is a particular choice that we make and is known as the M\o ller-Plesset type of partitioning. There are other choices that one can make for the energy-denominator, however, these do not necessarily have superior convergence properties over the simple M\o ller-Plesset partitioning \cite{Surjan04}.

Note that Eq. (\ref{it_eq_wfpt}) requires the matrix elements of the perturbation operator, $W$. A convenient definition of the $W$ operator is to split the initial Hamiltonian $H$ into two pieces, namely $H=H_0+W$. We define $H_0$ to be that part of the Hamiltonian operator that only connects many-body basis states that lie in the space $\nmax=0-10$. In other words, $H_0$ does not connect basis states from our reference space to the $\nmax=12$ space and satisfies the eigenvalue equation 
$H_0 | \Psi_{\rm ref} \rangle = \epsilon_{\rm ref}|\Psi_{\rm ref} \rangle$. The full Hamiltonian, $H$, does, however, connect basis states from $\nmax=12$ to the reference space. Thus, we can rewrite Eq. (\ref{it_eq_wfpt}), by replacing the $W$ operator with the full Hamiltonian, $H$, as follows.

\begin{equation}
\label{it_eq_wfptH}
| \psi_{\nmax=12,{\rm IT}}^{(1)} \rangle = |\Psi_{{\rm ref},\nmax=10}\rangle + \sum_{\nu \in \nmax=12} \frac{\langle \phi_{\nu} | H | \Psi_{{\rm ref},\nmax=10} \rangle} {\epsilon_{\nu}-\epsilon_{\rm ref,sp}}  | \phi_{\nu}  \rangle
\end{equation}

Such a form is extremely convenient since we do not need to calculate any other matrix elements than those we already have to calculate for the Hamiltonian operator. Equation (\ref{it_eq_wfptH}) indicates that the largest correction to the wavefunction is essentially determined by the amplitude of the corresponding $\nmax=12$ basis state. The amplitude of the basis state is, in turn, determined by the Hamiltonian matrix element between the $\nmax=12$ basis state and the reference state. This leads us to define the importance measure of a basis state, $\kappa_{\nu}$, to be

\begin{equation}
\label{it_eq_kappa}
\kappa_{\nu}=  \frac{|\langle \phi_{\nu} | H | \Psi_{{\rm ref},\nmax=10} \rangle|} {\epsilon_{\nu}-\epsilon_{\rm ref,sp}}.
\end{equation}

We can now use the importance measure as a way to set a threshold limit as to which basis states are included in the truncated $\nmax=12$ space. A typical value for $\kappa$ is on the order of a few $10^{-5}$. If we now set the threshold value of the importance measure to some value, say $3\times10^{-5}$, we only keep those basis states ($\phi_{\nu}$) in $\nmax=12$ for which $\kappa_{\nu}\geq 3\times10^{-5}$. Since some states have an importance measure lower than this threshold, we will discard those states, and thus, start to truncate the $\nmax=12$ space. In reality, the number of states discarded depends not only on the threshold value, but also on which $\nmax$ basis space is currently being evaluated. Typically, the largest $\nmax$ spaces are most heavily truncated, whereas in the first few $\nmax$ spaces most basis states are kept. This observation agrees with our intuitive notion that the components of the ground-state are dominated by basis states found in the lower oscillator shells.

\subsection{Properties of Importance-Truncation and {\em a Posteriori} Corrections}

The selection procedure of the many-body basis states that are kept in the importance-truncation calculation is based on a first-order perturbation theory result (for the wavefunction in $\nmax=12$). Using a particular value for $\kappa$ leads to a specific number of many-body basis states being kept that span the now incomplete $\nmax=12$ space. We diagonalize the Hamiltonian, $H$, in this incomplete $\nmax=12$ space. The diagonalization results in a ground-state energy, $E_{0,\kappa}^{(1)}$, associated with the truncated wavefunction, $|\Psi_{\nmax=12,\kappa}\rangle$. Note that this wavefunction, which results from the diagonalization of the Hamiltonian in the truncated $\nmax=12$ space, is not the same wavefunction $|\psi_{\nmax=12,\kappa}^{(1)}\rangle$, which is what we assume is a good approximation to the actual wavefunction as shown in Eq. (\ref{it_eq_wfptH}). Choosing a smaller value for $\kappa$ will result in more basis states being kept, a different truncated wavefunction, and thus will also result in a different ground-state energy. From the variational principle, we know that the calculated ground-state energy will decrease as we decrease the threshold value for $\kappa$. 

It is also possible to estimate the energy contribution of the discarded states by using second-order perturbation theory to determine the energy correction. The first-order energy correction vanishes, since we have defined the perturbation operator as $W=H-H_0$.

\begin{equation}
E^{(1)}=\langle \Psi_{\rm ref} | W | \Psi_{\rm ref} \rangle = \epsilon_{\rm ref} - \epsilon_{\rm ref} = 0
\end{equation} 

In order to determine a non-zero quantity for the correction to the energy we need to evaluate the second-order correction to the energy, by summing over all the discarded basis states, as shown below.

\begin{equation}
\label{it_eq_ptE2}
E_{0,\kappa}^{(2)}=-\sum_{ \substack{\nu = {\rm discarded} \\ \nmax=12}} \frac{|\langle \phi_{\nu} | H |\Psi_{{\rm ref},\nmax=10} \rangle |^2}{\epsilon_{\nu}-\epsilon_{\rm ref,sp}}
\end{equation}

Using this result, we can improve on the ground-state energy calculated in the truncated space, $E_{0,\kappa}^{(1)}$ by adding $E_{0,\kappa}^{(2)}$ to it. The resulting energy, 

\begin{equation}
\label{it_eq_ptE12}
E_{0,\kappa}^{(1+2)}=E_{0,\kappa}^{(1)} + E_{0,\kappa}^{(2)},
\end{equation}
has a smaller dependence on $\kappa$ than $E_{0,\kappa}^{(1)}$ does, as will be illustrated later. However, $E_{0,\kappa}^{(1+2)}$ is no longer variational, as it is constructed from a quantity $E_{0,\kappa}^{(2)}$, which is not a result obtained from the diagonalization of the Hamiltonian.

\subsection{The extension to excited states}

So far, we have only discussed targeting basis states in the larger $\nmax$ space from a reference state that we took as the ground-state in a smaller $\nmax$ space. We can easily extend the basis selection procedure for excited states by replacing the reference state with the desired excited state. In definite terms, this means that we define a $\kappa$ threshold value for each state in which we are interested, and evaluate all basis states in the larger $\nmax$ space for each reference state. The $\kappa$ threshold value is taken as the same numerical quantity in each case. Returning to our previous example where we selected the basis states for $\nmax=12$ from an $\nmax=10$ reference state, we now define $\kappa_{\nu}^{(m)}$ as the corresponding $\kappa$ for each of the $m$ (ground- and excited-) states present. Note that the energy denominator remains the same for all the reference states used. 

\begin{equation}
\label{it_eq_kappaEx}
\kappa_{\nu}^{(m)} = \frac{|\langle \phi_{\nu} | H | \Psi_{{\rm ref},\nmax=10}^{(m)} \rangle|} {\epsilon_{\nu}-\epsilon_{\rm ref,sp}}.
\end{equation}
The number of basis states kept per fixed value of $\kappa$ is larger for multiple reference states compared to the number kept when only one reference state is used. This is expected since the structure of higher-lying states might be quite different to the ground-state. The overlap between different reference states with the next-larger $\nmax$ basis states, $\langle \phi_{\nu} | H | \Psi_{{\rm ref},\nmax=10}^{(m)} \rangle$, will be very different depending on the structure of the relevant reference state.

\subsection{Implementation of Importance-Truncation}
\label{sec_it_implementation}
The Importance-Truncation procedure has been built into the {\em No-Core Shell Model Slater Determinant Code} ({\em NCSD}) \cite{ncsd}, which is a multi-processor code that has its roots in the Iowa State {\em Many-Fermion Dynamics (MFDn)} code \cite{mfd92,mfd94,mfdna,Maris10}. However, the current {\em NCSD} code differs substantially from {\em MFDn}. In particular, the code implements a hash table to look up which Slater determinants connect to each other under the action of various operators, such as the Hamiltonian. The relative simplicity of {\em NCSD} allows for an easy modification of the code, so that the importance-truncation selection procedure could be done. We will now describe how this is done in our code, specifically for the reference state being the ground-state (the extension to excited states is much the same). 

The reader might have formed the impression that one needs to specify only one value of $\kappa$, in order to do ``the calculation''. Although it is certainly possible to calculate a good approximation to the next larger $\nmax$ space wavefunction by using just one value for $\kappa$, it is not sufficient to determine the actual energy that the complete $\nmax$ space would give. One needs to perform several calculations for various $\kappa$ threshold values, each resulting in a specific ground-state energy, $E_{0,\kappa}$, so that an extrapolation to $\kappa=0$ can be performed on the $E_{0,\kappa}$ values. The extrapolation to $E_{0,\kappa=0}$, which will be discussed in great detail in Sec. \ref{sec_extrapolations}, yields what we assume is the true ground-state energy of the next-larger NCSM $\nmax$ space. Whether or not this is the case will also be addressed in Sec. \ref{sec_extrapolations}, in which we compare extrapolated IT energies to the full-space energies obtained from the {\em ANTOINE} code \cite{antoine}. 

In order to calculate the series of $E_{0,\kappa}$ values that we need, an efficient algorithm was developed as follows. At the start of every calculation we determine the smallest value of $\kappa$ we would like to use. Most often, we choose the minimum value to be $\kappa_{\rm min}=1.0\times10^{-5}$. Next, we construct {\em all} the basis states of the next-larger $\nmax$ space and save those to a master file on disk. The master file is split up according to how many processors are used for the IT-NCSM calculation, typically ranging from 768 to 1536 processors. Each processor reads in the list of unique basis states ($\phi_{\nu}$) assigned to it and determines through one Lanczos iteration, which basis states satisfy the requirement that $\kappa_{\nu}\geq \kappa_{\rm min}$. Those basis states that do satisfy this requirement are saved to a new file along with the calculated value of $\kappa_{\nu}$. This new file holds only a small fraction of the initial $\nmax$ many-body basis states.

Since we now have a list of all basis states that satisfy  $\kappa_{\nu}\geq \kappa_{\rm min}$, as well as their corresponding value for $\kappa_{\nu}$, we are in a position to perform a series of calculations, in which we now vary $\kappa$. We define a series of $\kappa$ values, for example, $\kappa= \{ 3.0,2.0.1.0 \}\times10^{-5}$, and begin the calculation at the largest $\kappa$ value. All states that now satisfy $\kappa_{\nu} \geq 3.0\times10^{-5}$ are read in from the saved file and are added to the many-body basis states already present. The resulting $E_{0,\kappa=3.0}$ energy is saved. The process repeats, in which we now add all the basis states that satisfy $\kappa_{\nu} \geq 2.0\times10^{-5}$ that were not previously added. This procedure is repeated until we have calculated all the $E_{0,\kappa}$ for all the values of $\kappa$ given. The resulting series of $E_{0,\kappa}$ values are then used to extrapolate to $E_{0,\kappa=0}$.

The above procedure has been specific for one $\nmax$ space. In our calculations, we employ a bootstrapping idea in which we apply importance-truncation to several $\nmax$ spaces in a sequential order. This is very similar to the IT-NCSM(seq) technique of Roth \cite{Roth09}. We choose to begin with a complete $\nmax=4$ space, from which we construct the truncated basis in $\nmax=6$ using the appropriate reference state. We then perform a series of calculations, as described above, in order to determine enough $E_{0,\kappa}$ values for the $\nmax=6$ space, so that a reasonable extrapolation can be made to $E_{0,\kappa=0}$. Once we have calculated the energy for the smallest chosen $\kappa$ value, we use that resulting wavefunction,
$| \Psi_{{\rm ref},\nmax=6,\kappa_{\rm min}} \rangle$, as the reference state for evaluating the $\nmax=8$ basis states. Besides checking {\em all} the basis states in $\nmax=8$, we also re-open the master list of {\em all} previously discarded basis states for $\nmax=6$, and check if any of those states are {\em now} kept. This point will be discussed along with other observations on the importance-truncation procedure in Sec. \ref{sec_observations}. When we calculate the energy contribution from the discarded basis states, we re-evaluate the contribution from all the states that are still discarded. For example, once we evaluate our $\nmax=8$ basis states, we calculate the energy contribution of the discarded states from $\nmax=8$ states as well as those that are still discarded in $\nmax=6$. 

\begin{eqnarray}
E_{0,\kappa}^{(2)}= &-& \sum_{\substack{\nu = {\rm discarded} \\ \nmax=8}} \frac{|\langle \phi_{\nu} | H |\Psi_{{\rm ref},\nmax=6} \rangle |^2}{\epsilon_{\nu}-\epsilon_{\rm ref,sp}} \nonumber \\
 &-&\sum_{\substack{\nu = {\rm discarded} \\ \nmax=6}} \frac{|\langle \phi_{\nu} | H |\Psi_{{\rm ref},\nmax=6} \rangle |^2}{\epsilon_{\nu}-\epsilon_{\rm ref,sp}}. 
\end{eqnarray}
This series of truncated $\nmax$ calculations continues until the desired $\nmax$ space is reached, which in all of our calculations is $\nmax=14$. To summarize, note that at the end of each series of calculations in a given $\nmax$ space, the wavefunction corresponding to the smallest $\kappa$ value is used as the reference state for evaluating the basis states in the next larger $\nmax$ space, since that wavefunction is the best approximation (for the specified $\kappa$) to the complete $\nmax$ space. For each $\nmax$ space, we re-evaluate all the basis states that have been discarded in the lower $\nmax$ spaces to check if any of them now satisfy the minimum $\kappa$ threshold.

\subsection{A comparison to the IT-NCSM calculations of Roth}

The implementation of IT-NCSM, which we described above, differs slightly from the implementation of the IT-NCSM(seq) of Roth in Ref. \cite{Roth09}. Two differences arise: 1) the order in which the calculations are performed and 2) the use of an additional truncation on the reference wavefunction $|\Psi_{{\rm ref},\nmax}\rangle$. We will briefly describe the differences between the two implementations. For a thorough discussion we refer readers to Ref. \cite{Kruse-thesis}.

We have already described how we proceed with our calculations in the previous subsection. In the case of the Roth calculations, one value of $\kappa$ is specified, followed by a calculation of the ground-state energies in sequential $\nmax$ spaces. Once the desired $\nmax$ space is reached, a new calculation is performed with a different value of $\kappa$. Our method differs in the sense that we start with the largest value of $\kappa$ and proceed to calculate the ground-state energies as a function of $\kappa$ in a fixed-$\nmax$ space, before moving onto the next $\nmax$ space. 

The additional truncation on the reference wavefunction is often referred to as the $c_{\rm min}$ cut, in which components of the reference wavefunction, roughly 10 times larger than $\kappa_{\rm min}$, are kept (smaller components are discarded). This truncated reference wavefunction is then used as the actual reference wavefunction for evaluating basis states in the next-larger $\nmax$ space. The implementation that we have used in this article never uses the $c_{\rm min}$ cut.

\section{Extrapolating in a given $\nmax$ space}
\label{sec_extrapolations}

In Sec. \ref{sec_it_implementation}, we pointed out that several values of $\kappa$ are used; each resulting in a ground-state energy, $E_{0,\kappa}$. These are then used to extrapolate to $E_{0,\kappa=0}$. At $\kappa=0$ in an IT-NCSM calculation, then all basis states are kept; thus, by extrapolating to $\kappa=0$, we hope to recover the ground-state energy of the complete $\nmax$ calculation. In this section, we will carefully analyze the extrapolation procedure and make a reasonable estimate of the uncertainty produced simply by using various extrapolation techniques. Such an analysis is new and needs to be done. We will show that different conclusions can be drawn from the extrapolations, depending upon how they were performed. It is not surprising to expect an uncertainty to be present in IT-NCSM calculations; however, we must point out that such results should be interpreted with care. One instance where some care should be exercised is in the extrapolation of a few $\nmax$ calculations to the infinite space ($\nmax=\infty$). We will demonstrate that each calculated IT-NCSM $\nmax$ ground-state energy is associated with a small but finite uncertainty. These uncertainties tend to grow as $\nmax$ increases. If an extrapolation to $\nmax=\infty$ is now performed on these $\nmax$ points, one should expect an uncertainty to be associated with the predicted infinite result. The uncertainty on the infinite result is influenced by the finite $\nmax$ IT-NCSM calculations and their respective uncertainties. This naturally leads to the question, {\em how large is the associated uncertainty in the $\nmax=\infty$ result?}

\begin{figure}
	\centering
		\includegraphics{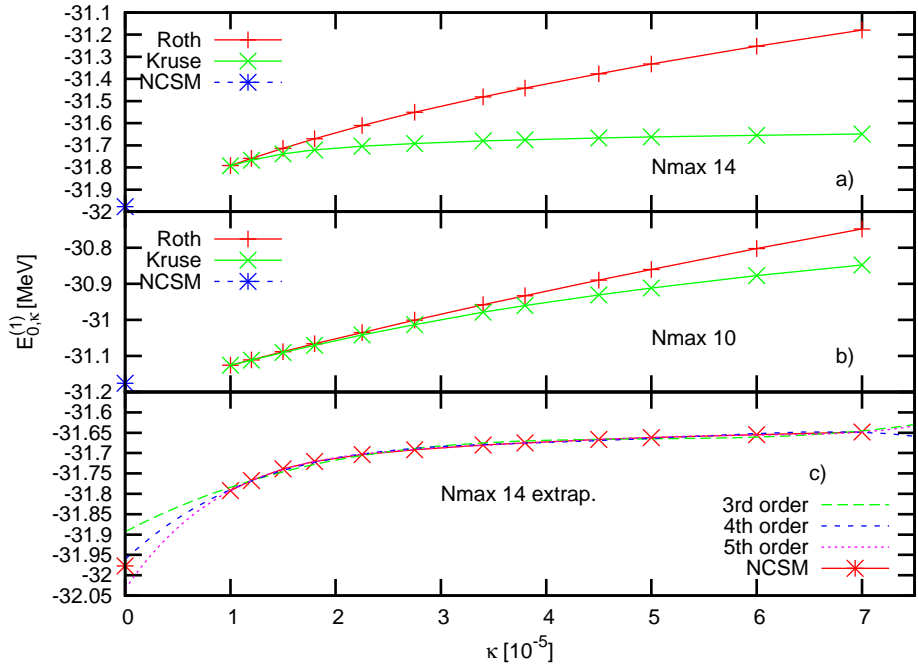}
	\caption{(color online) IT-NCSM calculated ground-state energies of $^6$Li in an a) $\nmax=14$ space and b) $\nmax=10$ space. The SRG-N3LO potential with a momentum-cutoff $\lambda=2.02$~fm$^{-1}$ as well as $\hbar\Omega=16$~MeV was used. The top two panels show the difference in calculated ground-state energies, when the Roth or present implementation is used. The NCSM result is shown at $\kappa=0$. The lower panel c) shows three different polynomial extrapolations to $E_{0,\kappa=0}$ using the $\nmax=14$ ground-state energies as calculated by our implementation. Note that the extrapolated values are different and are spread across a range of about 150 keV. }
	\label{fig:Li6-3fits}
\end{figure}

In Fig. \ref{fig:Li6-3fits}, we present a series of importance-truncated-calculated ground-state energies for $^6$Li in an $\nmax=14$ space, using a range of $\kappa$ values. The top two panels show the difference between the Roth implementation and that used here in the $\nmax=14$ and $\nmax=10$ space. We note that the difference in the nature of these curves can be understood from the underlying basis states present in the truncated space. Consider the $\nmax=14$ space as an example; and specifically the ground-state energy at $\kappa=7.0\times10^{-5}$. In the case of our implementation, one has kept all the basis states from $\nmax=6-12$ at $\kappa=1.0\times10^{-5}$, whereas in the Roth case one has only kept the basis states from $\nmax=6-12$ that satisfy $\kappa\geq7.0\times10^{-5}$.

We have chosen 12 values of $\kappa$ given by the set $\kappa~=~\{ 7.0,6.0.5.0,4.5,3.8,3.4,2.75,2.25,1.8,1.5,1.2,1.0 \} \times 10^{-5}$ which we will refer to as $\kappa-$grid points. This choice is arbitrary, although we did space the smallest $\kappa$ values closer together, since we intuitively know that the smallest $\kappa$ values have a larger effect on the extrapolation than the larger $\kappa$ values do. Our chosen range of $\kappa$, spanning from $\kappa=7.0-1.0 \times 10^{-5}$, is also to some extent arbitrary (one simply needs enough points to extrapolate to the ground-state energy at $\kappa=0$). Choosing too narrow a range could potentially affect the extrapolations in an undesired way by not capturing the general trend of the calculated $\kappa-$specific ground-state energies. Choosing too large a range, could bias the fitted functions towards $E_{0,\kappa}$ values that are associated with large values of $\kappa$. This brings us to our first point: The extrapolated values will depend on the chosen {\em range} of $\kappa$ and will also be influenced by the {\em spacing} of the values of $\kappa$. We will address these issues in Sec. \ref{subsec_kappa}.

Another look at Fig. \ref{fig:Li6-3fits} c) suggests that the extrapolated values will also depend on the chosen function that is to be extrapolated. In Fig. \ref{fig:Li6-3fits} c), we present three possible choices; a 3$^{rd}$, 4$^{th}$ and 5$^{th}$ order polynomial. The predicted ground-state energy in the $\nmax=14$ space has a range of about 150 keV between the 3$^{rd}$ order and 5$^{th}$ order polynomial. {\em A priori}, there is no class of functions that should be used in the extrapolations to the full-space result. Furthermore, we cannot make use of the Hellman-Feynman theorem, since the Hamiltonian does not explicitly depend on $\kappa$. In other words, there is no strict requirement that the extrapolated function should have a zero-derivative at $\kappa=0$. This brings us to our second point: The extrapolated values also depend on the {\em type} of function used to perform the extrapolation. An analysis of various functions will be presented in Sec. \ref{subsec_functions}. 

\begin{figure}
	\centering
		\includegraphics{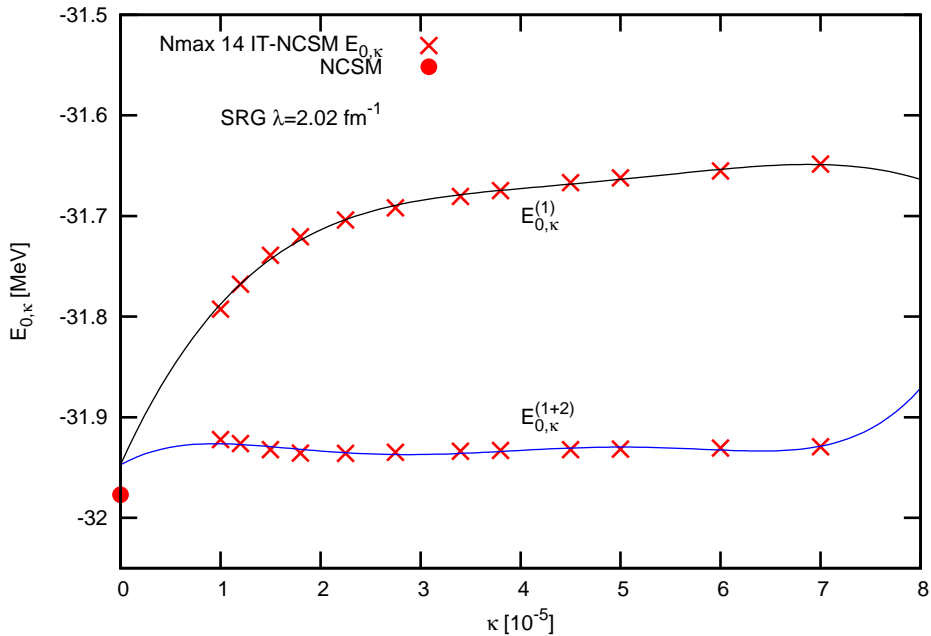}
	\caption{(color online) A constrained fit on both the first-,$E_{0,\kappa}^{(1)}$ (upper curve), and second-order energies, $E_{0,\kappa}^{(1+2)}$ (lower curve), using a 5$^{th}$ order polynomial. The NCSM parameters as well as the  $\kappa$-grid are the same as in Fig.\ref{fig:Li6-3fits}.}
	\label{fig:Li6-nmax14-2ndfit-pent}
\end{figure}

The results presented in Fig. \ref{fig:Li6-3fits} are variational, as they are calculated from a certain number of basis states, which in turn are determined from the importance-truncation selection procedure for a given value of $\kappa$, in which the Hamiltonian is diagonalized. An alternative way to fit the importance-truncated energies is to make use of the second-order corrections to the energy, $E_{0,\kappa}^{(2)}$, as shown in Eq. (\ref{it_eq_ptE2}). Formally, we know that as $\kappa \rightarrow 0$, both $E_{0,\kappa}^{(1)}$ (as shown in Fig. \ref{fig:Li6-3fits}) and $E_{0,\kappa}^{(1+2)}$ should meet at the same extrapolated point. It is, thus, also possible to do a constrained extrapolation of these two curves, one involving only the first-order energies $E_{0,\kappa}^{(1)}$, the other including the second-order corrections, $E_{0,\kappa}^{(1+2)}$, in such a way that both curves meet at the same point when $\kappa=0$. Such an extrapolation is shown in Fig. \ref{fig:Li6-nmax14-2ndfit-pent}, using the same NCSM parameters as in Fig. \ref{fig:Li6-3fits}. Although we have only shown the fit for the 5$^{th}$ order polynomial, it should be noted that using another polynomial will lead to a different extrapolated result. As will be shown later, when the constrained fit is used the spread in extrapolated ground-state energies is lower than that suggested in Fig. \ref{fig:Li6-3fits}. However, a spread in the extrapolated values does remain, and we would like to characterize how large that spread is. 


\subsection{Minimizing the effect on the chosen set of $\kappa$ values}
\label{subsec_kappa}

In Sec. \ref{sec_extrapolations}, we pointed out that the extrapolations to $E_{0,\kappa=0}$ depends on the chosen set of $\kappa$ values. In particular, the extrapolation depends on the {\em range} of the set, the {\em number} of $\kappa$-grid points, as well as their {\em spacing}. In this section, we analyze the dependence on the extrapolated ground-state energy on these quantities. Such an analysis is quite interesting for the following reason: A different choice of $\kappa$-grid points leads to a different extrapolated value of the ground-state energy. Typically the range of $\kappa$ is similar, spanning from a minimal value of a few $10^{-5}$ to a maximal value of about $20\times 10^{-5}$. In the larger $\nmax$ calculations, especially for the p-shell, it is computationally expensive to have $\kappa < 1.0\times10^{-5}$, since the number of states grows exponentially when the value of $\kappa$ is decreased.

Usually one fits the 12 points, shown in Fig. \ref{fig:Li6-3fits} and Fig. \ref{fig:Li6-nmax14-2ndfit-pent}, by using some specified low-order polynomial \cite{Roth09}. This, however, leads to one value of $E_{0,\kappa=0}$, without the ability to determine any uncertainty that is due solely to the extrapolation itself. One can have a feel for the uncertainty in the extrapolation by dropping the ground-state energy associated with the smallest value of $\kappa$ and then re-fitting the remaining points. The difference between these two extrapolations usually gives an initial estimate of the uncertainty. 

\begin{figure}
	\centering
		\includegraphics{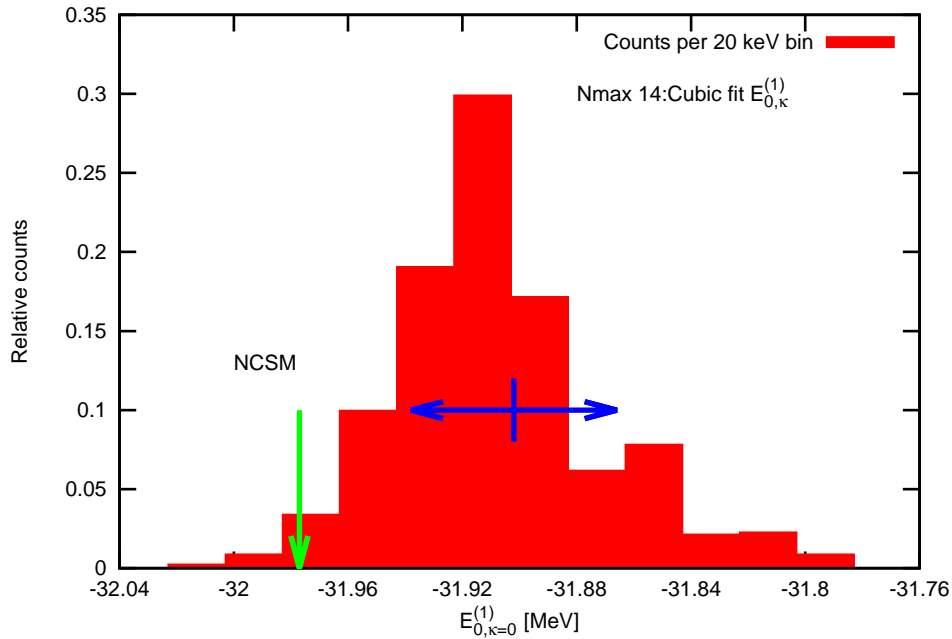}
	\caption{(color online) The normalized distribution of extrapolated ground-state energies of $^6$Li in the $\nmax=14$ space using $\hb=16$ MeV. We use the chiral NN N3LO interaction, softened by SRG to $\lambda=2.02$ fm$^{-1}$. The extrapolations are done using a cubic polynomial fitted only to the first-order energies, as was done in Fig. \ref{fig:Li6-3fits}. The extrapolated values are binned by 20 keV. We determine the median to be $E_{0,\kappa=0}=-31.902$ MeV and the standard deviation to be $\sigma_{{12}\choose{7}}=36$ keV as shown by the horizontal blue line. The green vertical line indicates the NCSM result.}
	\label{fig:Li6-1st-histo-cub7}
\end{figure}

We would like to provide an improved method for determining the uncertainty of IT-NCSM extrapolations. As a first estimate of the uncertainty produced by varying ranges and spacings of the grid-points, as well as the number of grid points, we use the following procedure. We begin by choosing all possible combinations of 7 out of our 12 available points, and for each of these ${{12}\choose{7}}=792$ sets we fit an extrapolating function to the data set and determine the extrapolated ground-state energy, $E_{0,\kappa=0}$. An example of the distribution of extrapolated energies using a cubic polynomial fitted to the first-order energies, $E_{0,\kappa}^{(1)}$, is shown in Fig. \ref{fig:Li6-1st-histo-cub7}. After calculating all the extrapolated ground-state energies that result from the 792 combinations of grid-points, we bin the results in 20 keV bins. From the distribution we calculate the median as well as the standard deviation (indicated by the blue horizontal line in Fig. \ref{fig:Li6-1st-histo-cub7}). We chose the median, instead of the average, as it is a statistical quantity that is not sensitive to outliers in the distribution. In the case of Fig. \ref{fig:Li6-1st-histo-cub7}, we determine the median to be $E_{0,\kappa=0}^{{12}\choose{7}}=-31.902$ MeV. The standard deviation of the distribution shown in Fig. \ref{fig:Li6-1st-histo-cub7} is $\sigma_{{12}\choose{7}}=36$ keV. Note that the area enclosed by one-standard deviation below and above the median value in the distribution, shown in Fig. \ref{fig:Li6-1st-histo-cub7}, encompasses roughly two-thirds of the extrapolated energies.

The preceding paragraph lays the foundations of our uncertainty analysis. We repeat the above procedure for another 3 sets of data, created from choosing ${12}\choose{8}$, ${12}\choose{9}$ and finally ${12}\choose{10}$ combinations of $\kappa$-grid points. For each data set we determine the median of the distribution, as shown in Fig. \ref{fig:Li6-1st-histo-cub7}, as well as the standard deviation. The median for each data set varies by at most a few keV, whereas the standard deviation decreases as the number of combinations of $\kappa$-grid points decreases. We determine our extrapolated ground-state energy, as well as the associated uncertainty in the extrapolation from the grid points by averaging over the calculated medians and standard deviations of the 4 data sets, using

\begin{eqnarray}
\label{eq_meansEn}
E_{0,\kappa=0}=\frac{E_{0,\kappa=0}^{{12}\choose{7}} + E_{0,\kappa=0}^{{12}\choose{8}} + E_{0,\kappa=0}^{{12}\choose{9}}  + E_{0,\kappa=0}^{{12}\choose{10}}}{4}. \\
\label{eq_meansStd}
\sigma = \frac{\sigma_{{12}\choose{7}} + \sigma_{{12}\choose{8}} + \sigma_{{12}\choose{9}} + \sigma_{{12}\choose{10}}}{4}.
\end{eqnarray}

We should point out that our determination of the uncertainty which is generated from various combinations of $\kappa-$grid points with which we associate $\sigma$, is only a first attempt at determining the potential uncertainty of the extrapolations. The calculated standard deviation will, in general, differ if fewer (or more) data sets are used in Eq. (\ref{eq_meansStd}). The important point that we want to make is that, although the uncertainty might change depending on the number of data sets used, the order of magnitude of the uncertainty, whether it be a few or tens of keV's will not change. Such an estimate does have implications, when extrapolations to $\nmax=\infty$ are performed. We also note that the standard deviation is generally smaller, when one uses the constrained extrapolations, as is shown in Fig. \ref{fig:Li6-nmax14-2ndfit-pent}. 

\subsection{Polynomial extrapolating functions}
\label{subsec_functions}

In Section \ref{subsec_kappa},  we focused our attention on determining the uncertainty that is generated from various combinations of $\kappa-$grid points. The objective of that section was to average over many different possible choices of grid-point configurations. In this section, we address the choice of extrapolation function. As we had mentioned in the introduction to Section \ref{sec_extrapolations}, there is no {\em a priori} justification to using one function over another. One simply goes by whether the chosen function, once fitted to the calculated $E_{0,\kappa}$, lies on top of the data or not. As can be seen from Fig. \ref{fig:Li6-3fits}, various ground-state energies are predicted, depending upon which function was chosen for the extrapolation. We will, thus, investigate various options that one might consider in fitting IT-NCSM calculated energies. We will use three different polynomials, a cubic, quartic, as well as a 5$^{th}$-order polynomial.

For each selected function, we repeat the procedure outlined in Sec. \ref{subsec_kappa}. Besides fitting the first-order results, $E_{0,\kappa}^{(1)}$, as in Fig. \ref{fig:Li6-3fits}, we also repeat the extrapolations using the same function for both the first- and second-order results, $E_{0,\kappa}^{(1)}$ and $E_{0,\kappa}^{(1+2)}$, respectively, as shown in Fig. \ref{fig:Li6-nmax14-2ndfit-pent}. Note that the constrained fits lead to a smaller standard deviation in the extrapolated ground-state energy.

\subsection{Estimates of extrapolation uncertainties}
\label{subsec_errorsuncertainties}

We will now present our calculated uncertainty estimates on the extrapolated ground-state energy in $^6$Li for the model spaces $\nmax=6-14$. Recall that the oscillator value is $\hbar\Omega=16$ MeV. The extrapolated ground-state energies are calculated as well as the standard deviation, which we associate with the uncertainty generated from the extrapolation, as explained in Section \ref{subsec_kappa}. We also present the results for using various extrapolating functions. The extrapolated results are compared to the NCSM ground-state energies, as shown in Table \ref{tab:hw16-gs1st} and \ref{tab:hw16-gs2nd}. 

\begin{table}
	\centering
		\begin{tabular}{|c|cc|cccccc|c|}
		\hline
		$\nmax$ & NCSM & IT-NCSM & Cubic [MeV] & $\sigma$ [keV] & Quartic [MeV] & $\sigma$ [keV]& 
		5$^{th}$-order [MeV] & $\sigma$ [keV] & NCSM [MeV] \\
		\hline
		6  & 0.198 & 0.162 & -28.601 &  $\approx$ 0 & -28.601 &  $\approx$ 0 & -28.601 & $\approx$ 0 & -28.602 \\
		8  & 1.579 & 1.077 &  -30.216 & 2   & -30.211 & 2   & -30.208 & 1    &  -30.213 \\
		10 & 9.693 & 3.291 & -31.207 & 2   & -31.204 & 3   & -31.197 & 4    & -31.176 \\
		12 & 48.888 & 6.487 & -31.714 & 10   & -31.744 & 6   & -31.741 & 21  & -31.713 \\
		14 & 211.286 & 9.544 & -31.899 & 25   & -31.971 & 33   & -32.046 & 29 & -31.977 \\
		\hline
		\end{tabular}
			\caption{The extrapolated ground-state energies in various $\nmax$ spaces for $^6$Li ($\lambda=2.02$ fm$^{-1}$) for $\hbar\Omega=16$~MeV, when using only the first-order IT-NCSM calculated points, $E_{0,\kappa}^{(1)}$. The table displays the various mean extrapolated values of the ground-state energy as well as the calculated standard deviations of the fits, which are indicated to the right of the corresponding extrapolated energy. The NCSM result, in which all basis states are kept, is shown in the right-most column. The dimension (in millions) of basis states are shown in the complete $\nmax$ space as well as in the importance-truncated space (column 2 and 3). Note that the basis in IT-NCSM is drastically reduced in the larger $\nmax$ values.}
	\label{tab:hw16-gs1st}
\end{table}

\begin{table}
	\centering
		\begin{tabular}{|c|cccccc|c|}
		\hline
		$\nmax$ & Cubic [MeV] & $\sigma$ [keV] & Quartic [MeV] & $\sigma$ [keV]& 
		5$^{th}$-order [MeV] & $\sigma$ [keV] & NCSM [MeV] \\
		\hline
		6  & -28.601 &  $\approx$ 0 & -28.602 &  1 & -28.602 &  2  & -28.602 \\
		8  & -30.217 & 2   & -30.211 & 2   & -30.208 & 1  & -30.213 \\
		10 & -31.194 & 1   & -31.196 & 1   & -31.195 & 2   & -31.176 \\
		12 & -31.685 & 6   & -31.702 & 5   & -31.712 & 2  & -31.713 \\
		14 & -31.902 & 9   & -31.925 & 12   & -31.952 & 13 & -31.977 \\
		\hline
		\end{tabular}
			\caption{The extrapolated ground-state energies in various $\nmax$ spaces for $^6$Li ($\lambda=2.02$ fm$^{-1}$) for $\hbar\Omega=16$~MeV, using the second-order IT-NCSM calculated points, $E_{0,\kappa}^{(1+2)}$. The table displays the various mean extrapolated values of the ground-state energy as well as the calculated standard deviations of the fits, which are indicated to the right of the corresponding extrapolated energy. The NCSM result, in which all basis states are kept, is shown in the right-most column.}
	\label{tab:hw16-gs2nd}
\end{table}

\begin{figure}
	\centering
		\includegraphics{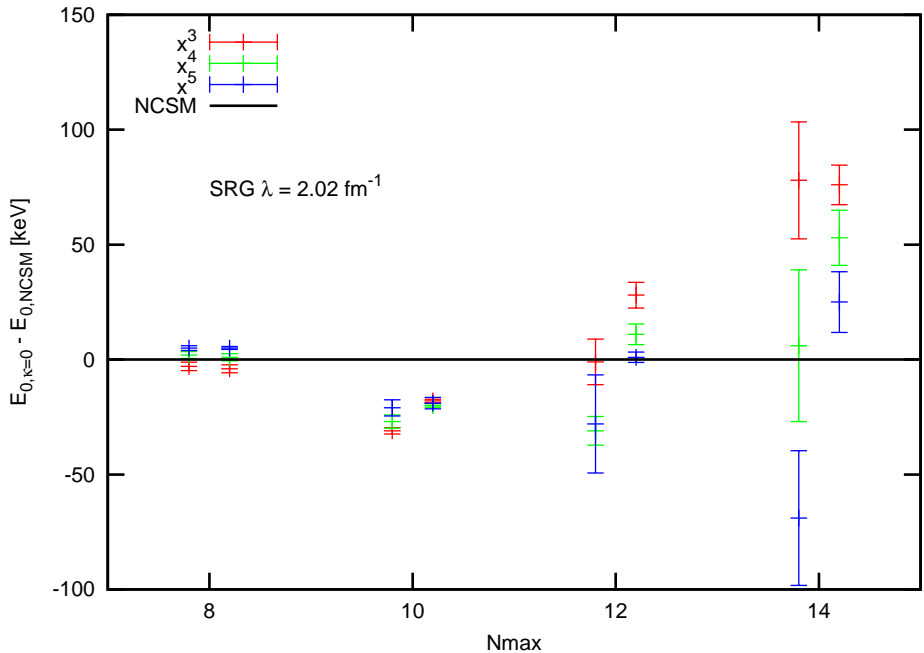}
	\caption{(color online) The plot shows the extrapolated ground-state energy relative to the NCSM ground-state energy (solid line), as well as the uncertainty, $\sigma$, that we determine from variations in the $\kappa$-grid points. The points to the left are the extrapolations that are generated from fitting just the first-order set of data, $E_{0,\kappa}^{(1)}$. Those to the right are the extrapolations when fitting the second-order corrections, $E_{0,\kappa}^{(1+2)}$. The oscillator value is $\hbar\Omega=16$~MeV.}
	\label{fig:Li6-hw16-diff-exact}
\end{figure}

We observe the following trends, presented by Table \ref{tab:hw16-gs1st}, Table \ref{tab:hw16-gs2nd} and Fig. \ref{fig:Li6-hw16-diff-exact}. The extrapolations to the NCSM ground-state energy for $\nmax=6-8$ are very good, being within one keV of the NCSM result and independent of the function or method used. The agreement with the NCSM result is not surprising as most of the many-body basis states are kept in those $\nmax$ spaces. Next, we observe that the uncertainty, $\sigma$, increases as $\nmax$ increases, but that it is smaller in the larger $\nmax$ spaces for the constrained second-order fits ($E_{0,\kappa}^{(1+2)}$) than the uncertainty for the corresponding first-order fits. Note that the uncertainty stated here is from variations in the combinations of $\kappa$-grid points. This result is also expected as now many basis states are discarded for the $\nmax=12-14$ spaces. However, note that for a given $\nmax$ space that the uncertainties associated with each extrapolating function are roughly the same. This indicates that at least at some level, choosing one function over another, does not necessarily decrease the uncertainty from variations in the $\kappa$-grid points. 

We also note that the mean extrapolated ground-state energy for a given $\nmax$ space can be quite different for various functions, varying as much as 50 keV for the $\nmax=14$ space, when either the first- ($E_{0,\kappa}^{(1)}$) or constrained second-order ($E_{0,\kappa}^{(1+2)}$) results are fitted. The spread of extrapolated ground-state energies among the various chosen functions is usually quite a bit larger than the uncertainty associated with the variations in the $\kappa$-grid points. Realistically, one does not have the NCSM calculations on hand, otherwise there would be no need for IT-NCSM, thus, characterizing the spread of the mean extrapolated ground-state energy is conceptually challenging. To illustrate this point, consider the results for $\nmax=12$ in Table \ref{tab:hw16-gs1st}. The cubic polynomial extrapolates to the NCSM result to within a keV, yet the quartic and 5$^{th}$-order polynomial overestimate the result by about 30 keV, which is twice as large as the uncertainty from the variations in the $\kappa$-grid points. If one does not know the NCSM result, one cannot make a reasonable guess as to which functional extrapolation is the correct one to use. Furthermore, it should be clear from the tables that the uncertainty from variations in the $\kappa$-grid points are smaller than the spread associated with the use of different extrapolating functions. 

\subsection{Asymptotically-correct extrapolating functions}

The observant reader will notice that the polynomial fits do not necessarily have the correct asymptotic behavior at $\kappa\rightarrow\infty$. The fitting routine might determine the leading coefficient of the polynomial function to be negative which in turn would cause the extrapolating function to decrease at a large value of $\kappa$. One might propose several extrapolating functions that have the correct asymptotic for large $\kappa$ as a way to boost confidence in the fitting procedure. Examples of such fitting functions are

\begin{eqnarray}
\label{eq_nonLinFits}
f(\kappa) &=& a+c\exp(-b\kappa^2)\left(1 + d\kappa + f\kappa^2\right)\nonumber \\
f(\kappa) &=& a + c\left(1-\tanh(b\kappa)\right)\left(1 + d\kappa + f\kappa^2\right) \nonumber\\
f(\kappa) &=& a+c\left(1-{\rm erf}(b\kappa) \right)\left(1 + d\kappa + f\kappa^2\right) \nonumber\\
f(\kappa) &=& a + c\exp(-b\kappa)\left(1 + d\kappa + f\kappa^2\right),
\end{eqnarray}
in which $b,c,d$ and $f$ are treated as fit-parameters. We fix $a$ to the ground-state energy determined in the previous $\nmax$ space. For example, when we fit the points for the $\nmax=14$ space, we will fix $a$ to the ground-state energy calculated at $\kappa=1.0\times10^{-5}$ in the $\nmax=12$ space. Such a constraint is possible in our calculations. There are both advantages and disadvantages to using these functions. The advantage is that we have fewer fit parameters than the quartic or $5^{th}$ order polynomials and furthermore, we have a definite constraint on the asymptotic value of the energy for large $\kappa$. The disadvantage is that the extrapolating functions, shown in Eq. (\ref{eq_nonLinFits}), have non-linear fit parameters. It is also true that non-linear fit parameters are difficult to optimize globally and have the potential of leading to unstable fits, in which small changes in the starting values of the parameters lead to different final parameters being determined by the fitting routines. In our extrapolations, we typically use the parameters from previously determined (successful) extrapolations, in order to minimize the chance of an unstable fit.

\begin{figure}[htbp]
\begin{center}
\includegraphics[]{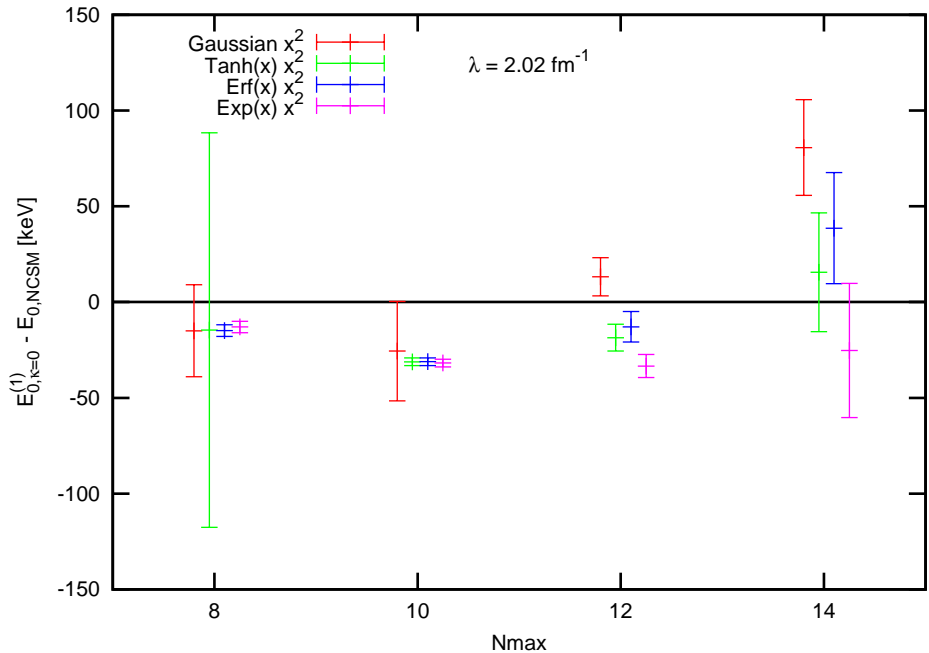}
\caption{(color online) The plot shows the extrapolated ground-state energy relative to the NCSM ground-state energy (solid line), as well as the uncertainty, $\sigma$, that we determine from variations in the $\kappa$-grid points. We have only fitted the first-order energies, $E_{0,\kappa}^{(1)}$, using the non-linear functions given in Eq. \ref{eq_nonLinFits}. The results show that the difference from the NCSM result as well as the spread in the extrapolated ground-state energy ($\sigma$) is about the same as for the polynomial functions, shown in Fig. \ref{fig:Li6-hw16-diff-exact}.}
\label{fig:nonLinFits}
\end{center}
\end{figure}

We repeat the extrapolations that were performed in Fig. \ref{fig:Li6-hw16-diff-exact}, using only the first-order calculated ground-state energies, $E_{0,\kappa}^{(1)}$. The result of the extrapolations is shown in Fig. \ref{fig:nonLinFits} for the extrapolating functions shown in Eq. \ref{eq_nonLinFits} and should be compared to Fig. \ref{fig:Li6-hw16-diff-exact}. We note that in some cases the calculated uncertainty ($\sigma$) is quite large, notably at $\nmax=8$.  At $\nmax=8$ the calculated energies, $E_{0,\kappa}^{(1)}$, form a convex function in $\kappa$. The extrapolating functions we have listed in Eq. \ref{eq_nonLinFits} are appropriate for concave functions as is typically the case for the calculated ground-state energies at $\nmax\geq10$ (see Fig. \ref{fig:Li6-3fits}). For the larger $\nmax$ spaces these functions seem to have a smaller uncertainty than the polynomial functions used earlier. This is particularly true for the $\nmax=14$ case in which the overall extent of the uncertainty is roughly 150 keV; the polynomial functions have an overall extent of about 200 keV. 

In the remaining part of the paper we will only use the polynomial functions to extrapolate to the ground-state energy in a given $\nmax$ space. This choice was made out of simplicity of fitting polynomial functions. Note that no inherent difficulties are present in fitting the asymptotically-correct extrapolation functions as shown in Eq. \ref{eq_nonLinFits}.

\section{Extrapolating to $\nmax=\infty$}
\label{sec_infty}

Often the final procedure in any NCSM calculation is to extrapolate the ground-state energies as a function of $\nmax$ to $\nmax=\infty$. The purpose of this procedure is to remove the model parameters $(\nmax,\hbar\Omega)$ from the calculations. The extrapolation to $\nmax=\infty$ removes the $\nmax$ dependence. The dependence on $\hbar\Omega$ is removed by the extrapolation to $\nmax=\infty$, because, in principle, when the complete basis is recovered, no $\hb$ dependence should remain. One usually chooses the HO frequency near the variational minimum of the ground-state (as we have done). This procedure was first used in \cite{Nav04}, in which the bare N3LO interaction was used to determine the ground-state energy of $^6$Li. We do, however, make the reader aware that a more satisfying extrapolation procedure is offered by considering concepts from EFT-theories (see \cite{Coon12, Furnstahl12} for an EFT-inspired approach to extrapolating NCSM ground-state energies).

In this work we will use the routinely used exponential extrapolation, in which the extrapolation to $\nmax=\infty$ is done by fitting the ground-state energy as a function of $\nmax$ to an exponential decay of the form $a\exp(-b\nmax)+c$. The constant $c$ represents the ground-state energy at $\nmax=\infty$. Furthermore, note that $b$ represents a fit-parameter and is not related to the oscillator-length in any way. In order to extrapolate the IT-NCSM ground-state energies to $\nmax=\infty$ one needs to consider a few points. These are the following: 1) the predicted ground-state energy various depending on the extrapolating function used, 2) there is an uncertainty in each of the extrapolating functions, 3) and there is no guarantee in general that any one function is better than any other function (for example, one can not choose the cubic polynomial extrapolations over the fifth-order extrapolations). To elaborate on the third point, it is also not clear that the fit involving say 10 $\kappa-$grid points should be better than the fit that used only 8 points (for a given function). 

\begin{figure}[htbp]
\begin{center}
\includegraphics[]{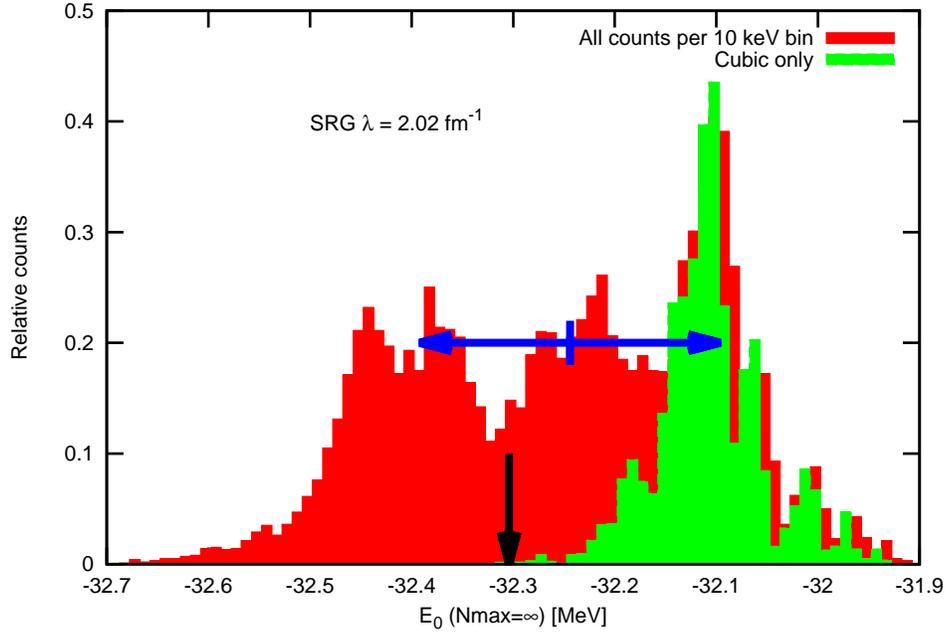}
\caption{(color online) The figure shows the distribution (in red) of extrapolated ground-state energies ($E_{0,\kappa=0}^{(1)}$) at $\nmax=\infty$ when an ensemble of 10 000 points is used as described in the text. The green distribution to the right represents the distribution that arises when only the cubic-polynomial extrapolated ground-state energies are considered. The blue arrow represents the uncertainty ($\sigma$) from the median value of the overall distribution whereas the black arrow represents the full NCSM extrapolated value. ($\hb=16$ MeV.)}
\label{fig:histoplotAll}
\end{center}
\end{figure}

We thus propose the following strategy in determining the uncertainty for the $\nmax=\infty$ extrapolations. At each $\nmax$ value we will randomly select one of the three polynomial functions that we have used as well as randomly select the number of $\kappa-$grid points used in determining the extrapolated ground-state energy for that specific $\nmax$ space. For example, we might randomly select the quartic-extrapolation that used 9 $\kappa-$grid points at $\nmax=8$, then select the cubic-extrapolation that used 7 $\kappa-$grid points for $\nmax=10$, and so on. This random selection is done for each $\nmax$ value between $\nmax=8-14$. Once we have selected these four ground-state energies, now as a function of $\nmax$, we extrapolate to $\nmax=\infty$ using the exponential decay function as described in the previous paragraph. The procedure is done for an ensemble of 10 000 randomly selected points; increasing this number does not change the ground-state energy or the uncertainty at $\nmax=\infty$. Fortunately, the resulting 10 000 extrapolations to $\nmax=\infty$ form a peaked distribution as shown in Fig. \ref{fig:histoplotAll}. In order to determine the final value of the ground-state energy at $\nmax=\infty$, we simply determine the median of the distribution as well as the standard deviation, as we have done before. Recall that we choose the median instead of the average as the median is insensitive to outliers in the distribution. 

\section{Observations on the Importance Truncation procedure}
\label{sec_observations}

In the previous section, we addressed the fundamentals of IT-NCSM calculations. In particular, we addressed the choice of the extrapolating function as well as the variation in $\kappa$-grid points. Here we address further questions, such as the dependence on the HO energy ($\hbar\Omega$), the SRG momentum-cutoff scale ($\lambda$), as well as the dependence of the IT-NCSM ground-state energies on the number of reference states used ({\em i.e.}, targeting excited states). These are discussed in Section \ref{sec_observations_hw},\ref{sec_observations_lambda} and \ref{sec_observations_ref}, respectively.

\subsection{The dependence on $\hbar\Omega$}
\label{sec_observations_hw}

\begin{figure}
	\centering
		\includegraphics{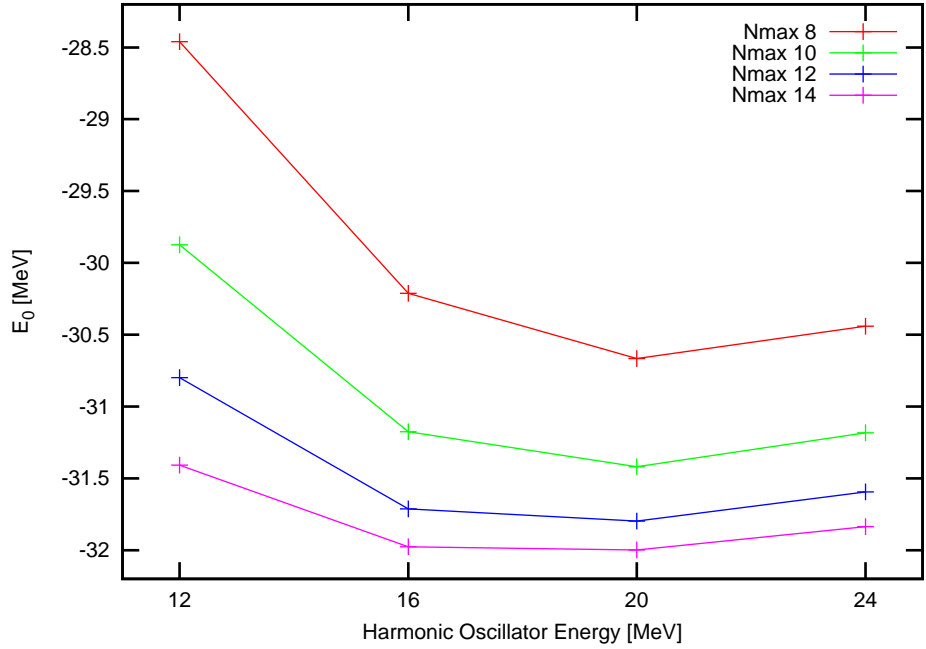}
	\caption{(color online) The HO energy ($\hbar\Omega$) dependence of NCSM ground-state energies for $^6$Li. The momentum-decoupling scale is $\lambda=2.02$ fm$^{-1}$. Note that $\hbar\Omega=16-20$ MeV corresponds to the optimal HO energy for this interaction. The solid lines are meant to guide the eye.}
	\label{fig:exact-all}
\end{figure}

All NCSM calculations have a dependence on the chosen HO energy, $\hbar\Omega$, even when bare interactions are used. However, the dependence on $\hbar\Omega$ can be minimized for a range of values. In practice, one typically chooses an $\hbar\Omega$ range resulting in the lowest ground-state energy of the largest $\nmax$ space employed. In Fig. \ref{fig:exact-all}, we plot the NCSM ground-state energies as a function of $\nmax$ for various HO energies (all basis states are kept). As $\nmax$ increases, the dependence on $\hbar\Omega$ decreases, leading to a range of possible values of the HO energy.

\begin{figure}
	\centering
		\includegraphics{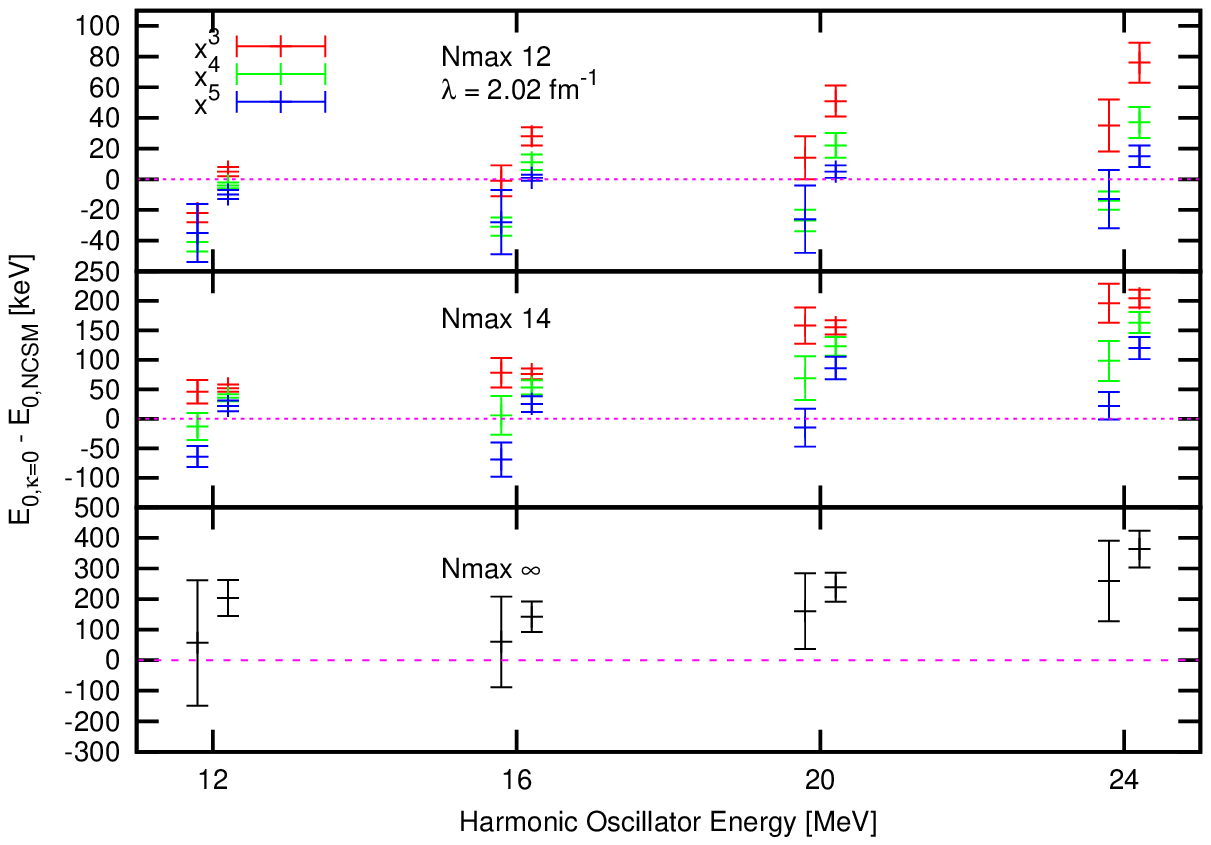}
	\caption{(color online) The figure shows the relative difference of the IT-NCSM extrapolated energies, $\Eone$ (left) and $\Etwo$ (right), to the NCSM ground-state energy (horizontal curve) as a function of the HO energy, $\hbar\Omega$. We also show the dependence on the various extrapolating functions ($\nmax=12-14$), indicating the uncertainty in the extrapolation technique. Note that there is a systematic drift away from the NCSM result, as $\hbar\Omega$ increases. The discrepancy increases as $\nmax$ increases. The solid black lines for the lowest panel indicate the uncertainty in the extrapolation to $\nmax=\infty$ as determined by the ensemble-averaged procedure of Sec. \ref{sec_infty}.}
	\label{fig:hw-multi-nmax12-14-inf}
\end{figure}

Having determined the NCSM results, we can now determine if the IT-NCSM extrapolated results show any dependence on $\hbar\Omega$. In other words, regardless of the $\hbar\Omega$ value used, do we extrapolate to the corresponding NCSM result, or is there a systematic difference as a function of $\hbar\Omega$? In Fig. \ref{fig:hw-multi-nmax12-14-inf}, we plot the difference of the extrapolated IT-NCSM ground-state energies, relative to the NCSM ground-state energy. Various extrapolating functions (cubic-, quartic- or 5$^{\rm th}$-order polynomials) are employed, using either the first- or second-order IT-NCSM extrapolated results, $\Eone$ or $\Etwo$, respectively. From  Fig. \ref{fig:hw-multi-nmax12-14-inf}, we can see that there is a systematic drift away from the NCSM result as $\hbar\Omega$ increases. The discrepancy also increases as $\nmax$ increases, averaging about 200 keV from the exact result for the $\nmax=\infty$ extrapolations. In Section \ref{sec_composition} we give some possible explanations for this type of behavior. 

\begin{table}
	\centering
		\begin{tabular}{|c|cc|cc|cc|cc|}
		\hline
		$\hb$ [MeV] & 12 & $\sigma$ [keV] & 16 & $\sigma$ [keV] & 20 & $\sigma$ [keV] & 24 & $\sigma$ [keV] \\
		\hline
		$E_{0,\kappa}^{(1)}$ [MeV] & -32.512 & 205 & -32.244 & 148 & -32.041 & 124 & -31.881 & 132 \\
		$E_{0,\kappa}^{(1+2)}$ [MeV] & -32.364 & 59 & -32.162 & 50 & -31.963 & 48 & -31.777 & 60 \\
		\hline
		NCSM [MeV] & -32.568 & - & -32.304 & - & -32.202 & - & -32.140 & - \\
		\hline
		\end{tabular}
			\caption{The table shows the extrapolated ground-state energy at $\nmax=\infty$ as a function of the harmonic oscillator energy $\hb$. The extrapolations as well as the uncertainties are determined by the procedure outlined in Sec. \ref{sec_infty}. We show both the first- as well as second-order extrapolations; these are the rows labeled $E_{0,\kappa}^{(1)}$ and $E_{0,\kappa}^{(1+2)}$, respectively. The comparison to a NCSM extrapolation can be graphically seen in the lower panel of Fig. \ref{fig:hw-multi-nmax12-14-inf}.}
			\label{tab:hw-nmax-infty-202}
\end{table}

\subsection{The dependence on the SRG momentum-decoupling scale ($\lambda$)}
\label{sec_observations_lambda}

We now investigate the dependence on the SRG momentum-decoupling scale, $\lambda$. The NN chiral EFT N3LO potential, evolved to a momentum-decoupling scale of $\lambda=$ 1.5 fm$^{-1}$, has recently been used in the NCSM/RGM calculation of $d-\alpha$ scattering \cite{Nav11-da}. At this value of $\lambda$, the off-shell characteristics of the two-body potential have been changed in such a way as to have the effect of producing a binding energy for $^6$Li similar to that obtained including the chiral EFT NNN N2LO potential (the two-body terms are retained up to N3LO). This behavior is due to the non-unitarity of the SRG procedure, when only two-body terms are kept in the RG evolution. The effect for $^6$Li is demonstrated in \cite{Jurg11} (see Fig. 11). It is, therefore, of interest to compare two different SRG evolved potentials, in order to see if the importance truncation selection procedure behaves differently in the two cases.



\begin{figure}
	\centering
		\includegraphics{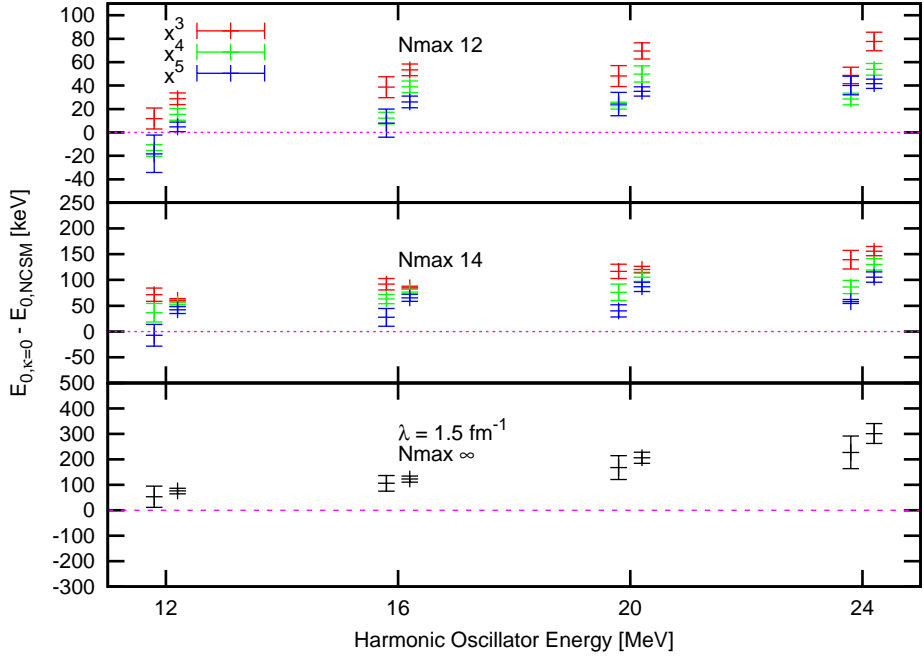}
	\caption{(color online) The figure shows the relative difference of the IT-NCSM extrapolated energies, $\Eone$ (left) and $\Etwo$ (right), to the  NCSM ground-state energy (horizontal curve), as a function of the HO energy, $\hbar\Omega$. We also show the dependence on the various extrapolating functions, indicating the uncertainty in the extrapolation technique. Note that there is a systematic drift away from the NCSM result as $\hbar\Omega$ increases. A similar observation was made in Fig. \ref{fig:hw-multi-nmax12-14-inf}. Note that we have used the same scale as in Fig. \ref{fig:hw-multi-nmax12-14-inf}, so that the two figures may easily be compared.}
	\label{fig:multi-nmax12-14-inf-150}
\end{figure}

\begin{figure}
	\centering
		\includegraphics{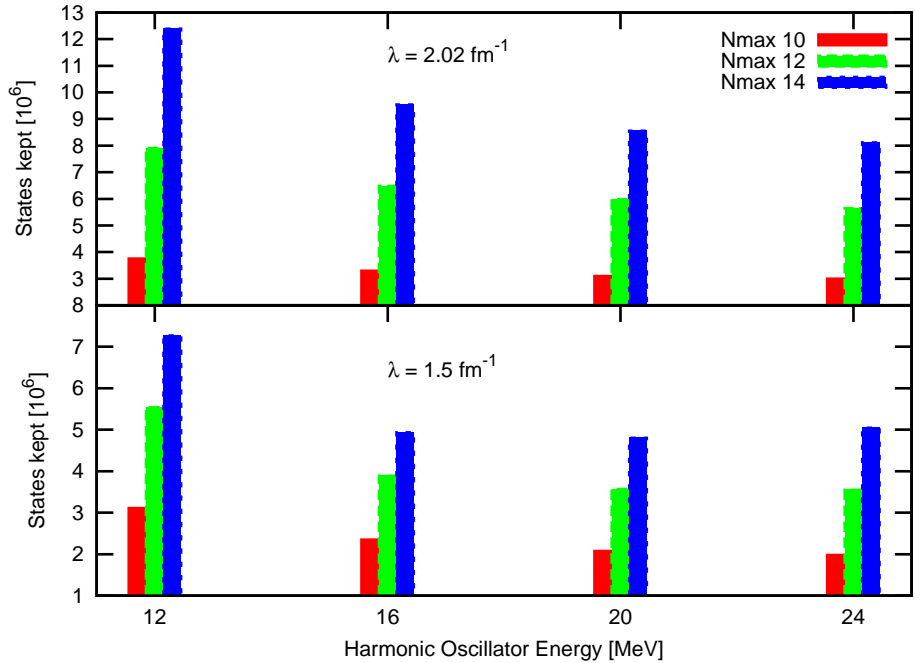}
	\caption{(color online) The number of basis states kept, as a function of $\nmax$, for the $\lambda={\textrm{2.02 fm}^{-1}}$ (top) and $\lambda={\textrm{1.5 fm}^{-1}}$ (bottom) SRG evolved N3LO interaction. Note that for the softer interaction, $\lambda={\textrm{1.5  fm}^{-1}}$, IT-NCSM keeps fewer basis states from the larger $\nmax$ spaces. We also note that as $\hbar\Omega$ increases, fewer basis states are kept for the larger $\nmax$ spaces. }
	\label{fig:dims-multi}
\end{figure}

In Fig. \ref{fig:multi-nmax12-14-inf-150}, we plot the extrapolated IT-NCSM results as a function of the HO energy, $\hbar\Omega$. This figure should be compared to Fig. \ref{fig:hw-multi-nmax12-14-inf} where we set $\lambda={\textrm{2.02 fm}^{-1}}$. Note that the trends are similar, but the relative uncertainty to the NCSM results is a bit smaller for $\lambda={\textrm{1.5 fm}^{-1}}$. The lower-$\lambda$ interactions are much softer, therefore, the convergence in $\nmax$ is much quicker. In this case, the IT-NCSM procedure selects fewer basis states for $\lambda={\textrm{1.5 fm}^{-1}}$ than it does for $\lambda={\textrm{2.02 fm}^{-1}}$, since for the softer potentials, fewer high-lying $\nmax$ basis states are required to reach convergence. This point is clearly illustrated in Fig. \ref{fig:dims-multi} where we plot the number of basis states kept, as a function of $\nmax$ for both types of SRG evolved potentials. 

\subsection{Further comments on the $\hbar\Omega$ dependence}
\label{sec_composition}

Figure \ref{fig:dims-multi} shows another interesting trend; fewer basis states are kept as $\hbar\Omega$ increases. In particular, note that the basis dimension is only about $\frac{3}{4}$ of the size for $\hbar\Omega=24$ MeV than it is for $\hbar\Omega=12$ MeV. 


In Fig. \ref{fig:hw-multi-nmax12-14-inf} and Fig. \ref{fig:multi-nmax12-14-inf-150}, we showed that the IT-NCSM extrapolated results shift {\em away} from the NCSM results as $\hbar\Omega$ increases. The explanation lies in the definition of the importance measure $\kappa$. Recall that $\kappa$ is inversely proportional to $\hbar\Omega$. Thus, for $\hbar\Omega$=24 MeV, the matrix elements $|\langle \phi_{\nu} | H | \Psi_{\rm ref} \rangle|$ would have to be twice as large as they are for $\hbar\Omega$=12 MeV, in order for $|\phi_{\nu}\rangle$ to be kept as a basis state. The matrix element itself also depends on $\hbar\Omega$, but this dependence must be weaker than the linear dependence in the denominator of $\kappa$ (since fewer states are kept as $\nmax$ increases). Using this argument, it is not too surpizing that the IT-NCSM results would have a dependence on the HO energy and that, in general, the IT-NCSM results would be less reliable for larger values of $\hbar\Omega$ (if the same minimum value of $\kappa$ is used), than for smaller values. We should also point out that since the IT-NCSM basis is an incomplete $\nmax$ space ({\em i.e.}, truncated), we no longer have a complete decomposition of center-of-mass from intrinsic states. Thus, a small amount of center of mass contamination is expected, which would increase as $\hbar\Omega$ increases. This issue has been addressed in \cite{Roth09b}.

\subsection{Using multiple reference states}
\label{sec_observations_ref}

One final feature we would like to investigate is the behavior of the excited-state spectrum in IT-NCSM calculations. In order to reliably calculate the excited states of an IT-NCSM calculation, one needs to employ a reference state for each state that is to be calculated. In our case, we desire to calculate the ground-state ($J^\pi=1^+$) and the first two excited states, corresponding to $J^\pi=3^+,0^+$. We use as initial reference states from $\nmax=4$ each of the three states to generate the basis states that are kept in $\nmax=6$. Since we are now using several reference states, the basis tends to be larger than if only one reference state is used. Thus, we have used a different set of importance-measure grid points, $\kappa=\{7.00,6.00,5.00,4.00,3.75,3.50,3.25,3.00,2.75,2.50,2.25,2.00\}\times10^{-5}$. This range is a bit smaller than the one we previously used for a single reference state and only extends to $\kappa=2.0\times10^{-5}$ instead of $\kappa=1.0\times10^{-5}$.  

\begin{figure}
	\centering
		\includegraphics{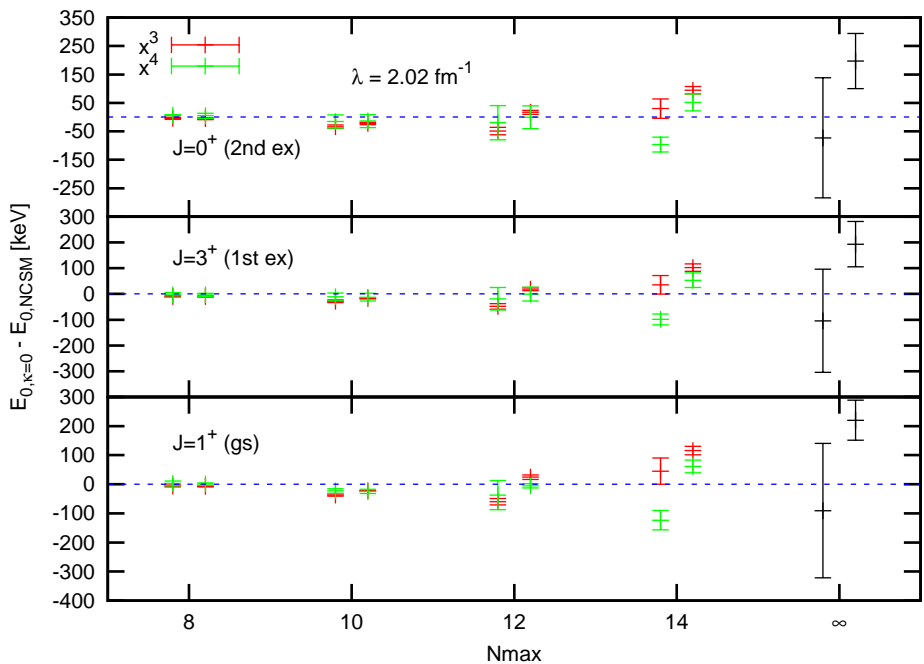}
	\caption{(color online) The figure shows the relative difference to the NCSM result (horizontal line), for various extrapolating techniques, $\Eone$ (left) or $\Etwo$ (right), as a function of increasing $\nmax$. In this case three reference states were used ($J^\pi=1^+,3^+,0^+)$. We calculate the ground-state and the first two excited states. Note that the overall trend is the same for all three states, indicating that IT-NCSM performs equally well for excited states as for the ground-state. We have also performed the extrapolation to $\nmax=\infty$ (black lines) by using the ensemble-averaging technique described in Sec. \ref{sec_infty}. ($\hbar\Omega=16$ MeV).}
	\label{fig:shifted-hw16ex-all}
\end{figure}

In Fig. \ref{fig:shifted-hw16ex-all}, we plot the relative difference to the NCSM result (horizontal line), for various extrapolating techniques, $\Eone$ (left) or $\Etwo$ (right), as a function of increasing $\nmax$ ($\hbar\Omega=16$~MeV). This is to be compared with Fig. \ref{fig:Li6-hw16-diff-exact}. The overall trend is the same between the two plots, indicating that the difference between IT-NCSM and NCSM calculations does not increase for the excited states, and that the difference is, in general, the same size as before (about 100 keV for $\nmax=14$). In other words, the excitation spectrum can be calculated with the same degree of accuracy as for the ground-state. The extrapolation to $\nmax=\infty$ has a larger uncertainty than what has been shown in the earlier figures. The uncertainty is especially large for the extrapolation that uses the first-order energies ($E_{0,\kappa=0}^{(1)}$). Considering only the first-order extrapolation, one sees that the predictions of the cubic- or quartic-polynomial extrapolation either under- or over-estimate the NCSM result as $\nmax$ increases. Thus, the result is that the ensemble-averaging procedure at $\nmax=\infty$ has a large uncertainty since the two polynomials essentially force the $\nmax=\infty$ extrapolation in opposite directions (compared to the NCSM).

\begin{figure}
	\centering
		\includegraphics{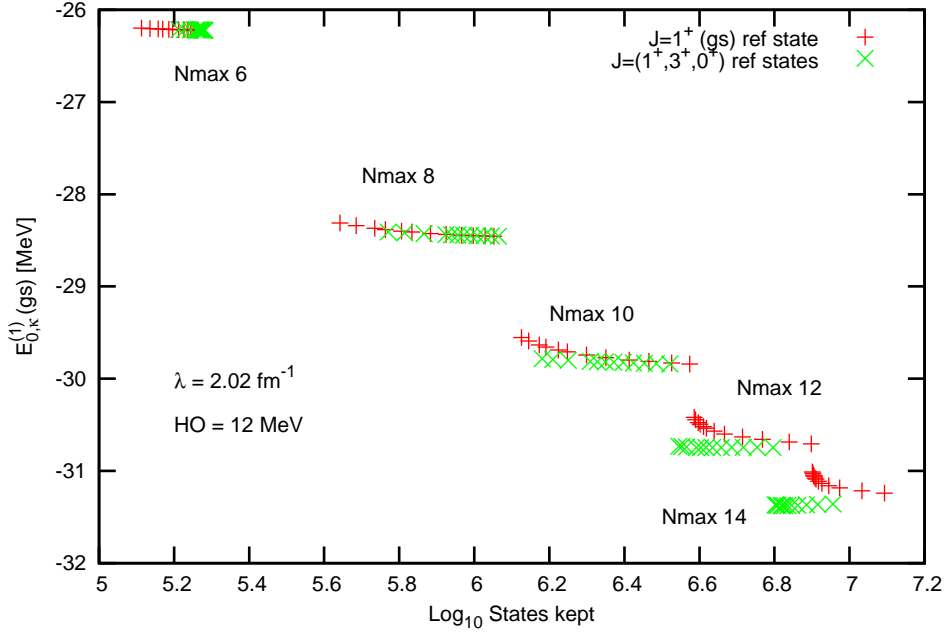}
	\caption{(color online) The figure shows the ground-state energy at $\hbar\Omega=12$ MeV, as a function of the Logarithm (base 10) of the number of basis states kept. The curve with (+) signs corresponds to when only the ground-state is used as a reference state, wheres the curve marked ($\times$) shows the behavior, when the lowest three states are used as reference states. Note that these energies correspond only to the first-order results, $E_{0,\kappa}^{(1)}$, and that the two sets of kappa grid points are not identical. The figure shows that basis states coming from higher $\nmax$ spaces  significantly lower the ground-state energy (note the decrease in the ground-state energy between each $\nmax$ space) as opposed to basis states in the same $\nmax$ space that have a smaller value of $\kappa$. Furthermore, using several reference states leads to an approximately constant dependence for the ground-state energy on the number of basis states kept. }
	\label{fig:hw12-dims-E0}
\end{figure}

An interesting comparison to make between using one or several reference states is to determine the behavior of the ground-state energy as a function of the number of basis states kept. This analysis is shown in Fig. \ref{fig:hw12-dims-E0}, for $\hbar\Omega=12$ MeV. We chose that particular value of $\hbar\Omega$, because the largest number of basis states are kept for this case. From the figure, one can deduce two interesting points. The first is that the additional reference states are selecting basis states that were previously not selected, as can be seen at the start of each $\nmax$ space. These additional basis states tend to make the functional dependence of the ground-state energy as a function of the size of the basis approximately constant. It can also be seen that when multiple reference states are used, fewer states are needed than before in order to achieve the same ground-state energy as with a single reference state. The second point has to do with the lowering of the ground-state energy as a function of $\nmax$. Note that higher $\nmax$ contributions significantly lower the ground-state energy, when compared to simply adding more states in a single $\nmax$ space. In other words, note that the drop in energy between $\nmax=12\rightarrow14$ is larger than the drop in energy in just the $\nmax=12$ space, which results from adding all the basis states that are kept. Such a feature could hold promise for doing configuration-interaction calculations, in which one- and two particle-hole excitations are created on top of a Hartree-Fock state determined in a small $\nmax$ space. Most of the binding energy is gained from adding the most significant configurations found in larger $\nmax$ spaces. However, we did notice in our IT-NCSM calculations that as $\nmax$ increases, previously discarded basis states in the lower $\nmax$ spaces {\em do} become relevant at some stage and are added back into the basis by our basis evaluation procedure. This also explains why we typically use all states up to and including $\nmax=4$; those basis states will be added to the basis in any case by IT-NCSM. In fact, by the time we have completed our $\nmax=14$ calculation, almost all of the $\nmax=6$ states have been added to the basis, even though initially a fair number of those states were discarded at the start of the calculation. Such behavior makes sense, since we expect low-lying states to have components mostly found in the lower oscillator shells. 

We expect many of the same results to hold for multiple reference states, as shown for the single reference state calculations. In particular, we have determined that the $\hbar\Omega$ dependence, as shown in Fig. \ref{fig:hw-multi-nmax12-14-inf}, still persists, when multiple reference states are used. This once again leads us to the conclusion that the dependence stems from $\kappa$ being inversely proportional to $\hbar\Omega$.

\section{Conclusion \label{sec_conclusions}}

We have presented a detailed investigation of IT-NCSM calculations for $^6$Li in which we have studied the dependence of IT-NCSM on various parameters. These include the behavior of IT-NCSM as a function of the model space $\nmax$, the HO energy $\hbar\Omega$, the extrapolating functions used for the two types of data sets ($E_{0,\kappa}^{(1)}$ or $E_{0,\kappa}^{(1+2)}$), the SRG momentum-decoupling scale $\lambda$, as well as the influence on the basis selection procedure, when multiple reference states are used. The IT-NCSM calculations were then compared to NCSM calculations, as a way to estimate the efficiency of the procedure. We find that the extrapolations used in IT-NCSM, using either the first- or second-order results ($\Eone$ or $\Etwo$), give similar results, even when different extrapolating functions are used. At $\nmax=14$ we find that the IT-NCSM extrapolated ground-state energies differ from the NCSM ground-state energies by about 100-150~keV, whereas in the extrapolation to $\nmax=\infty$ the difference is about 250~keV. The IT-NCSM calculations show no quantitative difference for the two SRG momentum-decoupling scales that we used ($\lambda=2.02$~fm$^{-1}$ and $\lambda=1.50$~fm$^{-1}$) in terms of the extrapolation uncertainties or in regard to the difference to the full NCSM calculations.
Two new features were seen in these calculations that have not been reported before: 1.) IT-NCSM calculations seem to deteriorate in quality as $\hbar\Omega$ increases, when the same $\kappa-$grid is used (see Fig. \ref{fig:Li6-hw16-diff-exact} and Fig. \ref{fig:multi-nmax12-14-inf-150}); 2.) using several reference states leads to a better basis selection for the ground-state energy, than using just a single reference state (see Fig. \ref{fig:hw12-dims-E0}). We propose that future IT-NCSM calculations should use multiple reference states to select the basis states, use smaller $\kappa$ threshold limits as $\hbar\Omega$ increases and provide a reasonable uncertainty estimate of the $\nmax=\infty$ extrapolated energies.

In this work, we have tried to provide an improved analysis of uncertainties in IT-NCSM calculations. However, we must point out that the only way we are able to be confident in our IT-NCSM calculations is by direct comparison to full NCSM calculations. In situations where the NCSM calculations can be performed, IT-NCSM energy-spectra have compared favorably to NCSM \cite{Roth07,Roth09,PhysRevC.82.034609}. However, in cases where no NCSM calculations exist, IT-NCSM calculations should be treated carefully, especially when ground-state energies are extrapolated to $\nmax=\infty$. Recall that in order to extrapolate the ground-state energies to $\nmax=\infty$ two extrapolations must be performed. The first extrapolation is performed in a given $\nmax$ space to estimate the NCSM result ($\kappa=0$). As $\nmax$ increases the extrapolated ground-state energy typically underestimates the full NCSM ground-state energy, especially as $\hb$ increases for a fixed $\kappa-$grid. The underestimation of the ground-state energy in larger $\nmax$ spaces inadvertently drives the extrapolation to $\nmax=\infty$ to a smaller binding energy than what is predicted with the full NCSM. Furthermore, one is free to choose a low-order polynomial function (or asymptotically-correct function) to extrapolate the ground-state energies in each $\nmax$ space. Each functional choice leads to a different prediction and different uncertainty for the ground-state energy.  In our $\nmax=\infty$ extrapolation, we have taken into account the uncertainties of each $\nmax$ ground-state energy and have determined the uncertainty at $\nmax=\infty$ over a range of $\hb$ to be on the order of 100-200~keV for $\lambda=2.02$~fm$^{-1}$ and on the order of 50-100~keV for $\lambda=1.5$~fm$^{-1}$ (see Fig. \ref{fig:hw-multi-nmax12-14-inf} and Fig. \ref{fig:multi-nmax12-14-inf-150}, respectively). The smaller uncertainty for $\lambda=1.5$~fm$^{-1}$ is due to the softer nature of the underlying NN interaction.

\section{Acknowledgements}

M.K.G.K would like to thank S.A. Coon for many discussions regarding the results of this paper as well as providing critical readings of the manuscript. M.K.G.K would also like to thank D. Toussaint for discussions on estimating uncertainties. M.K.G.K. and B.R.B acknowledge partial support from the U.S. NSF grants PHY-0555396 and PHY-0854912. M.K.G.K, E.D.J and W.E.O acknowledge funding from the U. S. DOE/SC/NP. P.N. acknowledges support from the Natural Sciences and Engineering Research Council of Canada (NSERC) Grant No. 401945-2011. TRIUMF receives funding via a contribution through the National Research Council Canada. Numerical calculations have been performed at the LLNL LC facilities
supported by LLNL under Contract No. DE-AC52-07NA27344.

\bibliographystyle{apsrev} 
\bibliography{bib-nov30}

\begin{thebibliography}{55}
\expandafter\ifx\csname natexlab\endcsname\relax\def\natexlab#1{#1}\fi
\expandafter\ifx\csname bibnamefont\endcsname\relax
  \def\bibnamefont#1{#1}\fi
\expandafter\ifx\csname bibfnamefont\endcsname\relax
  \def\bibfnamefont#1{#1}\fi
\expandafter\ifx\csname citenamefont\endcsname\relax
  \def\citenamefont#1{#1}\fi
\expandafter\ifx\csname url\endcsname\relax
  \def\url#1{\texttt{#1}}\fi
\expandafter\ifx\csname urlprefix\endcsname\relax\def\urlprefix{URL }\fi
\providecommand{\bibinfo}[2]{#2}
\providecommand{\eprint}[2][]{\url{#2}}

\bibitem[{\citenamefont{Faddeev}(1960)}]{Faddeev60}
\bibinfo{author}{\bibfnamefont{L.~D.} \bibnamefont{Faddeev}},
  \bibinfo{journal}{Zh. Eksp. Teor. Fiz.} \textbf{\bibinfo{volume}{39}},
  \bibinfo{pages}{1459} (\bibinfo{year}{1960}).

\bibitem[{\citenamefont{Faddeev}(1961)}]{Faddeev61}
\bibinfo{author}{\bibfnamefont{L.~D.} \bibnamefont{Faddeev}},
  \bibinfo{journal}{Sov. Phys.-JETP} \textbf{\bibinfo{volume}{12}},
  \bibinfo{pages}{1014} (\bibinfo{year}{1961}).

\bibitem[{\citenamefont{Yakubovsky}(1967)}]{Yak67}
\bibinfo{author}{\bibfnamefont{O.~A.} \bibnamefont{Yakubovsky}},
  \bibinfo{journal}{Sov. J. Nucl. Phys.} \textbf{\bibinfo{volume}{5}},
  \bibinfo{pages}{937} (\bibinfo{year}{1967}).

\bibitem[{\citenamefont{Gl\"ockle and Kamada}(1993)}]{Glockle93}
\bibinfo{author}{\bibfnamefont{W.}~\bibnamefont{Gl\"ockle}} \bibnamefont{and}
  \bibinfo{author}{\bibfnamefont{H.}~\bibnamefont{Kamada}},
  \bibinfo{journal}{Phys. Rev. Lett.} \textbf{\bibinfo{volume}{71}},
  \bibinfo{pages}{971} (\bibinfo{year}{1993}),
  \urlprefix\url{http://link.aps.org/doi/10.1103/PhysRevLett.71.971}.

\bibitem[{\citenamefont{Ciesielski and Carbonell}(1998)}]{Cies}
\bibinfo{author}{\bibfnamefont{F.}~\bibnamefont{Ciesielski}} \bibnamefont{and}
  \bibinfo{author}{\bibfnamefont{J.}~\bibnamefont{Carbonell}},
  \bibinfo{journal}{Phys. Rev. C} \textbf{\bibinfo{volume}{58}},
  \bibinfo{pages}{58} (\bibinfo{year}{1998}),
  \urlprefix\url{http://link.aps.org/doi/10.1103/PhysRevC.58.58}.

\bibitem[{\citenamefont{Viviani et~al.}(1995)\citenamefont{Viviani, Kievsky,
  and Rosati}}]{Viviani}
\bibinfo{author}{\bibfnamefont{M.}~\bibnamefont{Viviani}},
  \bibinfo{author}{\bibfnamefont{A.}~\bibnamefont{Kievsky}}, \bibnamefont{and}
  \bibinfo{author}{\bibfnamefont{S.}~\bibnamefont{Rosati}},
  \bibinfo{journal}{Few-Body Systems} \textbf{\bibinfo{volume}{18}},
  \bibinfo{pages}{25} (\bibinfo{year}{1995}), ISSN \bibinfo{issn}{0177-7963},
  \bibinfo{note}{10.1007/s006010050002},
  \urlprefix\url{http://dx.doi.org/10.1007/s006010050002}.

\bibitem[{\citenamefont{Barnea et~al.}(1999)\citenamefont{Barnea, Leidemann,
  and Orlandini}}]{Barnea}
\bibinfo{author}{\bibfnamefont{N.}~\bibnamefont{Barnea}},
  \bibinfo{author}{\bibfnamefont{W.}~\bibnamefont{Leidemann}},
  \bibnamefont{and}
  \bibinfo{author}{\bibfnamefont{G.}~\bibnamefont{Orlandini}},
  \bibinfo{journal}{Nuclear Physics A} \textbf{\bibinfo{volume}{650}},
  \bibinfo{pages}{427 } (\bibinfo{year}{1999}), ISSN \bibinfo{issn}{0375-9474},
  \urlprefix\url{http://www.sciencedirect.com/science/article/pii/S037594749900113X}.

\bibitem[{\citenamefont{Navr\'atil
  et~al.}(2000{\natexlab{a}})\citenamefont{Navr\'atil,
  Kamuntavi\ifmmode~\check{c}\else \v{c}\fi{}ius, and Barrett}}]{Nav00Jac}
\bibinfo{author}{\bibfnamefont{P.}~\bibnamefont{Navr\'atil}},
  \bibinfo{author}{\bibfnamefont{G.~P.}
  \bibnamefont{Kamuntavi\ifmmode~\check{c}\else \v{c}\fi{}ius}},
  \bibnamefont{and} \bibinfo{author}{\bibfnamefont{B.~R.}
  \bibnamefont{Barrett}}, \bibinfo{journal}{Phys. Rev. C}
  \textbf{\bibinfo{volume}{61}}, \bibinfo{pages}{044001}
  (\bibinfo{year}{2000}{\natexlab{a}}),
  \urlprefix\url{http://link.aps.org/doi/10.1103/PhysRevC.61.044001}.

\bibitem[{\citenamefont{Navr\'atil and Barrett}(1998)}]{Nav98}
\bibinfo{author}{\bibfnamefont{P.}~\bibnamefont{Navr\'atil}} \bibnamefont{and}
  \bibinfo{author}{\bibfnamefont{B.~R.} \bibnamefont{Barrett}},
  \bibinfo{journal}{Phys. Rev. C} \textbf{\bibinfo{volume}{57}},
  \bibinfo{pages}{562} (\bibinfo{year}{1998}),
  \urlprefix\url{http://link.aps.org/doi/10.1103/PhysRevC.57.562}.

\bibitem[{\citenamefont{Navr\'atil
  et~al.}(2000{\natexlab{b}})\citenamefont{Navr\'atil, Vary, and
  Barrett}}]{Nav00}
\bibinfo{author}{\bibfnamefont{P.}~\bibnamefont{Navr\'atil}},
  \bibinfo{author}{\bibfnamefont{J.~P.} \bibnamefont{Vary}}, \bibnamefont{and}
  \bibinfo{author}{\bibfnamefont{B.~R.} \bibnamefont{Barrett}},
  \bibinfo{journal}{Phys. Rev. Lett.} \textbf{\bibinfo{volume}{84}},
  \bibinfo{pages}{5728} (\bibinfo{year}{2000}{\natexlab{b}}),
  \urlprefix\url{http://link.aps.org/doi/10.1103/PhysRevLett.84.5728}.

\bibitem[{\citenamefont{Navr\'atil et~al.}(2009)\citenamefont{Navr\'atil,
  Quaglioni, Stetcu, and Barrett}}]{NavRev09}
\bibinfo{author}{\bibfnamefont{P.}~\bibnamefont{Navr\'atil}},
  \bibinfo{author}{\bibfnamefont{S.}~\bibnamefont{Quaglioni}},
  \bibinfo{author}{\bibfnamefont{I.}~\bibnamefont{Stetcu}}, \bibnamefont{and}
  \bibinfo{author}{\bibfnamefont{B.~R.} \bibnamefont{Barrett}},
  \bibinfo{journal}{Journal of Physics G: Nuclear and Particle Physics}
  \textbf{\bibinfo{volume}{36}}, \bibinfo{pages}{083101}
  (\bibinfo{year}{2009}),
  \urlprefix\url{http://stacks.iop.org/0954-3899/36/i=8/a=083101}.

\bibitem[{\citenamefont{Pudliner et~al.}(1997)\citenamefont{Pudliner,
  Pandharipande, Carlson, Pieper, and Wiringa}}]{GFMC97}
\bibinfo{author}{\bibfnamefont{B.~S.} \bibnamefont{Pudliner}},
  \bibinfo{author}{\bibfnamefont{V.~R.} \bibnamefont{Pandharipande}},
  \bibinfo{author}{\bibfnamefont{J.}~\bibnamefont{Carlson}},
  \bibinfo{author}{\bibfnamefont{S.~C.} \bibnamefont{Pieper}},
  \bibnamefont{and} \bibinfo{author}{\bibfnamefont{R.~B.}
  \bibnamefont{Wiringa}}, \bibinfo{journal}{Phys. Rev. C}
  \textbf{\bibinfo{volume}{56}}, \bibinfo{pages}{1720} (\bibinfo{year}{1997}),
  \urlprefix\url{http://link.aps.org/doi/10.1103/PhysRevC.56.1720}.

\bibitem[{\citenamefont{Wiringa et~al.}(2000)\citenamefont{Wiringa, Pieper,
  Carlson, and Pandharipande}}]{GFMC00}
\bibinfo{author}{\bibfnamefont{R.~B.} \bibnamefont{Wiringa}},
  \bibinfo{author}{\bibfnamefont{S.~C.} \bibnamefont{Pieper}},
  \bibinfo{author}{\bibfnamefont{J.}~\bibnamefont{Carlson}}, \bibnamefont{and}
  \bibinfo{author}{\bibfnamefont{V.~R.} \bibnamefont{Pandharipande}},
  \bibinfo{journal}{Phys. Rev. C} \textbf{\bibinfo{volume}{62}},
  \bibinfo{pages}{014001} (\bibinfo{year}{2000}),
  \urlprefix\url{http://link.aps.org/doi/10.1103/PhysRevC.62.014001}.

\bibitem[{\citenamefont{Pieper et~al.}(2002)\citenamefont{Pieper, Varga, and
  Wiringa}}]{GFMC02}
\bibinfo{author}{\bibfnamefont{S.~C.} \bibnamefont{Pieper}},
  \bibinfo{author}{\bibfnamefont{K.}~\bibnamefont{Varga}}, \bibnamefont{and}
  \bibinfo{author}{\bibfnamefont{R.~B.} \bibnamefont{Wiringa}},
  \bibinfo{journal}{Phys. Rev. C} \textbf{\bibinfo{volume}{66}},
  \bibinfo{pages}{044310} (\bibinfo{year}{2002}),
  \urlprefix\url{http://link.aps.org/doi/10.1103/PhysRevC.66.044310}.

\bibitem[{\citenamefont{Pieper and Wiringa}(2001)}]{GFMCrev}
\bibinfo{author}{\bibfnamefont{S.~C.} \bibnamefont{Pieper}} \bibnamefont{and}
  \bibinfo{author}{\bibfnamefont{R.~B.} \bibnamefont{Wiringa}},
  \bibinfo{journal}{Annual Review of Nuclear and Particle Science}
  \textbf{\bibinfo{volume}{51}}, \bibinfo{pages}{53} (\bibinfo{year}{2001}),
  \urlprefix\url{http://www.annualreviews.org/doi/abs/10.1146/annurev.nucl.51.101701.132506}.

\bibitem[{\citenamefont{Bartlett and Musia\l{}}(2007)}]{Bartlett07}
\bibinfo{author}{\bibfnamefont{R.~J.} \bibnamefont{Bartlett}} \bibnamefont{and}
  \bibinfo{author}{\bibfnamefont{M.}~\bibnamefont{Musia\l{}}},
  \bibinfo{journal}{Rev. Mod. Phys.} \textbf{\bibinfo{volume}{79}},
  \bibinfo{pages}{291} (\bibinfo{year}{2007}),
  \urlprefix\url{http://link.aps.org/doi/10.1103/RevModPhys.79.291}.

\bibitem[{\citenamefont{Hagen et~al.}(2008)\citenamefont{Hagen, Papenbrock,
  Dean, and Hjorth-Jensen}}]{Hagen08}
\bibinfo{author}{\bibfnamefont{G.}~\bibnamefont{Hagen}},
  \bibinfo{author}{\bibfnamefont{T.}~\bibnamefont{Papenbrock}},
  \bibinfo{author}{\bibfnamefont{D.~J.} \bibnamefont{Dean}}, \bibnamefont{and}
  \bibinfo{author}{\bibfnamefont{M.}~\bibnamefont{Hjorth-Jensen}},
  \bibinfo{journal}{Phys. Rev. Lett.} \textbf{\bibinfo{volume}{101}},
  \bibinfo{pages}{092502} (\bibinfo{year}{2008}),
  \urlprefix\url{http://link.aps.org/doi/10.1103/PhysRevLett.101.092502}.

\bibitem[{\citenamefont{Navr\'atil et~al.}(2007)\citenamefont{Navr\'atil,
  Gueorguiev, Vary, Ormand, and Nogga}}]{Nav07}
\bibinfo{author}{\bibfnamefont{P.}~\bibnamefont{Navr\'atil}},
  \bibinfo{author}{\bibfnamefont{V.~G.} \bibnamefont{Gueorguiev}},
  \bibinfo{author}{\bibfnamefont{J.~P.} \bibnamefont{Vary}},
  \bibinfo{author}{\bibfnamefont{W.~E.} \bibnamefont{Ormand}},
  \bibnamefont{and} \bibinfo{author}{\bibfnamefont{A.}~\bibnamefont{Nogga}},
  \bibinfo{journal}{Phys. Rev. Lett.} \textbf{\bibinfo{volume}{99}},
  \bibinfo{pages}{042501} (\bibinfo{year}{2007}),
  \urlprefix\url{http://link.aps.org/doi/10.1103/PhysRevLett.99.042501}.

\bibitem[{\citenamefont{Maris et~al.}(2010{\natexlab{a}})\citenamefont{Maris,
  Shirokov, and Vary}}]{Maris10-F14}
\bibinfo{author}{\bibfnamefont{P.}~\bibnamefont{Maris}},
  \bibinfo{author}{\bibfnamefont{A.~M.} \bibnamefont{Shirokov}},
  \bibnamefont{and} \bibinfo{author}{\bibfnamefont{J.~P.} \bibnamefont{Vary}},
  \bibinfo{journal}{Phys. Rev. C} \textbf{\bibinfo{volume}{81}},
  \bibinfo{pages}{021301} (\bibinfo{year}{2010}{\natexlab{a}}),
  \urlprefix\url{http://link.aps.org/doi/10.1103/PhysRevC.81.021301}.

\bibitem[{\citenamefont{Goldberg et~al.}(2010)\citenamefont{Goldberg, Roeder,
  Rogachev, Chubarian, Johnson, Fu, Alharbi, Avila, Banu, McCleskey
  et~al.}}]{Goldberg10}
\bibinfo{author}{\bibfnamefont{V.}~\bibnamefont{Goldberg}},
  \bibinfo{author}{\bibfnamefont{B.}~\bibnamefont{Roeder}},
  \bibinfo{author}{\bibfnamefont{G.}~\bibnamefont{Rogachev}},
  \bibinfo{author}{\bibfnamefont{G.}~\bibnamefont{Chubarian}},
  \bibinfo{author}{\bibfnamefont{E.}~\bibnamefont{Johnson}},
  \bibinfo{author}{\bibfnamefont{C.}~\bibnamefont{Fu}},
  \bibinfo{author}{\bibfnamefont{A.}~\bibnamefont{Alharbi}},
  \bibinfo{author}{\bibfnamefont{M.}~\bibnamefont{Avila}},
  \bibinfo{author}{\bibfnamefont{A.}~\bibnamefont{Banu}},
  \bibinfo{author}{\bibfnamefont{M.}~\bibnamefont{McCleskey}},
  \bibnamefont{et~al.}, \bibinfo{journal}{Physics Letters B}
  \textbf{\bibinfo{volume}{692}}, \bibinfo{pages}{307 } (\bibinfo{year}{2010}),
  ISSN \bibinfo{issn}{0370-2693},
  \urlprefix\url{http://www.sciencedirect.com/science/article/pii/S0370269310009068}.

\bibitem[{\citenamefont{Entem and Machleidt}(2003)}]{EM500}
\bibinfo{author}{\bibfnamefont{D.~R.} \bibnamefont{Entem}} \bibnamefont{and}
  \bibinfo{author}{\bibfnamefont{R.}~\bibnamefont{Machleidt}},
  \bibinfo{journal}{Phys. Rev. C} \textbf{\bibinfo{volume}{68}},
  \bibinfo{pages}{041001} (\bibinfo{year}{2003}),
  \urlprefix\url{http://link.aps.org/doi/10.1103/PhysRevC.68.041001}.

\bibitem[{\citenamefont{Epelbaum et~al.}(2002)\citenamefont{Epelbaum, Nogga,
  Gl\"ockle, Kamada, Mei\ss{}ner, and Wita\l{}a}}]{Epelbaum02}
\bibinfo{author}{\bibfnamefont{E.}~\bibnamefont{Epelbaum}},
  \bibinfo{author}{\bibfnamefont{A.}~\bibnamefont{Nogga}},
  \bibinfo{author}{\bibfnamefont{W.}~\bibnamefont{Gl\"ockle}},
  \bibinfo{author}{\bibfnamefont{H.}~\bibnamefont{Kamada}},
  \bibinfo{author}{\bibfnamefont{U.-G.} \bibnamefont{Mei\ss{}ner}},
  \bibnamefont{and}
  \bibinfo{author}{\bibfnamefont{H.}~\bibnamefont{Wita\l{}a}},
  \bibinfo{journal}{Phys. Rev. C} \textbf{\bibinfo{volume}{66}},
  \bibinfo{pages}{064001} (\bibinfo{year}{2002}),
  \urlprefix\url{http://link.aps.org/doi/10.1103/PhysRevC.66.064001}.

\bibitem[{\citenamefont{Bogner et~al.}(2003{\natexlab{a}})\citenamefont{Bogner,
  Kuo, and Schwenk}}]{Bogner03a}
\bibinfo{author}{\bibfnamefont{S.}~\bibnamefont{Bogner}},
  \bibinfo{author}{\bibfnamefont{T.}~\bibnamefont{Kuo}}, \bibnamefont{and}
  \bibinfo{author}{\bibfnamefont{A.}~\bibnamefont{Schwenk}},
  \bibinfo{journal}{Physics Reports} \textbf{\bibinfo{volume}{386}},
  \bibinfo{pages}{1 } (\bibinfo{year}{2003}{\natexlab{a}}), ISSN
  \bibinfo{issn}{0370-1573},
  \urlprefix\url{http://www.sciencedirect.com/science/article/pii/S0370157303002953}.

\bibitem[{\citenamefont{Bogner et~al.}(2003{\natexlab{b}})\citenamefont{Bogner,
  Kuo, Schwenk, Entem, and Machleidt}}]{Bogner03b}
\bibinfo{author}{\bibfnamefont{S.}~\bibnamefont{Bogner}},
  \bibinfo{author}{\bibfnamefont{T.}~\bibnamefont{Kuo}},
  \bibinfo{author}{\bibfnamefont{A.}~\bibnamefont{Schwenk}},
  \bibinfo{author}{\bibfnamefont{D.}~\bibnamefont{Entem}}, \bibnamefont{and}
  \bibinfo{author}{\bibfnamefont{R.}~\bibnamefont{Machleidt}},
  \bibinfo{journal}{Physics Letters B} \textbf{\bibinfo{volume}{576}},
  \bibinfo{pages}{265 } (\bibinfo{year}{2003}{\natexlab{b}}), ISSN
  \bibinfo{issn}{0370-2693},
  \urlprefix\url{http://www.sciencedirect.com/science/article/pii/S0370269303015594}.

\bibitem[{\citenamefont{Bogner et~al.}(2007)\citenamefont{Bogner, Furnstahl,
  and Perry}}]{Bogner07}
\bibinfo{author}{\bibfnamefont{S.~K.} \bibnamefont{Bogner}},
  \bibinfo{author}{\bibfnamefont{R.~J.} \bibnamefont{Furnstahl}},
  \bibnamefont{and} \bibinfo{author}{\bibfnamefont{R.~J.} \bibnamefont{Perry}},
  \bibinfo{journal}{Phys. Rev. C} \textbf{\bibinfo{volume}{75}},
  \bibinfo{pages}{061001} (\bibinfo{year}{2007}),
  \urlprefix\url{http://link.aps.org/doi/10.1103/PhysRevC.75.061001}.

\bibitem[{\citenamefont{Jurgenson et~al.}(2009)\citenamefont{Jurgenson,
  Navr\'atil, and Furnstahl}}]{Jurg09}
\bibinfo{author}{\bibfnamefont{E.~D.} \bibnamefont{Jurgenson}},
  \bibinfo{author}{\bibfnamefont{P.}~\bibnamefont{Navr\'atil}},
  \bibnamefont{and} \bibinfo{author}{\bibfnamefont{R.~J.}
  \bibnamefont{Furnstahl}}, \bibinfo{journal}{Phys. Rev. Lett.}
  \textbf{\bibinfo{volume}{103}}, \bibinfo{pages}{082501}
  (\bibinfo{year}{2009}),
  \urlprefix\url{http://link.aps.org/doi/10.1103/PhysRevLett.103.082501}.

\bibitem[{\citenamefont{Jurgenson et~al.}(2011)\citenamefont{Jurgenson,
  Navr\'atil, and Furnstahl}}]{Jurg11}
\bibinfo{author}{\bibfnamefont{E.~D.} \bibnamefont{Jurgenson}},
  \bibinfo{author}{\bibfnamefont{P.}~\bibnamefont{Navr\'atil}},
  \bibnamefont{and} \bibinfo{author}{\bibfnamefont{R.~J.}
  \bibnamefont{Furnstahl}}, \bibinfo{journal}{Phys. Rev. C}
  \textbf{\bibinfo{volume}{83}}, \bibinfo{pages}{034301}
  (\bibinfo{year}{2011}),
  \urlprefix\url{http://link.aps.org/doi/10.1103/PhysRevC.83.034301}.

\bibitem[{\citenamefont{Bogner et~al.}(2010)\citenamefont{Bogner, Furnstahl,
  and Schwenk}}]{Bogner10}
\bibinfo{author}{\bibfnamefont{S.}~\bibnamefont{Bogner}},
  \bibinfo{author}{\bibfnamefont{R.}~\bibnamefont{Furnstahl}},
  \bibnamefont{and} \bibinfo{author}{\bibfnamefont{A.}~\bibnamefont{Schwenk}},
  \bibinfo{journal}{Progress in Particle and Nuclear Physics}
  \textbf{\bibinfo{volume}{65}}, \bibinfo{pages}{94 } (\bibinfo{year}{2010}),
  ISSN \bibinfo{issn}{0146-6410},
  \urlprefix\url{http://www.sciencedirect.com/science/article/pii/S0146641010000347}.

\bibitem[{\citenamefont{Lisetskiy et~al.}(2008)\citenamefont{Lisetskiy,
  Barrett, Kruse, Navratil, Stetcu, and Vary}}]{Lisetsky08}
\bibinfo{author}{\bibfnamefont{A.~F.} \bibnamefont{Lisetskiy}},
  \bibinfo{author}{\bibfnamefont{B.~R.} \bibnamefont{Barrett}},
  \bibinfo{author}{\bibfnamefont{M.~K.~G.} \bibnamefont{Kruse}},
  \bibinfo{author}{\bibfnamefont{P.}~\bibnamefont{Navratil}},
  \bibinfo{author}{\bibfnamefont{I.}~\bibnamefont{Stetcu}}, \bibnamefont{and}
  \bibinfo{author}{\bibfnamefont{J.~P.} \bibnamefont{Vary}},
  \bibinfo{journal}{Phys. Rev. C} \textbf{\bibinfo{volume}{78}},
  \bibinfo{pages}{044302} (\bibinfo{year}{2008}),
  \urlprefix\url{http://link.aps.org/doi/10.1103/PhysRevC.78.044302}.

\bibitem[{\citenamefont{Roth and Navr\'atil}(2007)}]{Roth07}
\bibinfo{author}{\bibfnamefont{R.}~\bibnamefont{Roth}} \bibnamefont{and}
  \bibinfo{author}{\bibfnamefont{P.}~\bibnamefont{Navr\'atil}},
  \bibinfo{journal}{Phys. Rev. Lett.} \textbf{\bibinfo{volume}{99}},
  \bibinfo{pages}{092501} (\bibinfo{year}{2007}),
  \urlprefix\url{http://link.aps.org/doi/10.1103/PhysRevLett.99.092501}.

\bibitem[{\citenamefont{Roth}(2009)}]{Roth09}
\bibinfo{author}{\bibfnamefont{R.}~\bibnamefont{Roth}}, \bibinfo{journal}{Phys.
  Rev. C} \textbf{\bibinfo{volume}{79}}, \bibinfo{pages}{064324}
  (\bibinfo{year}{2009}),
  \urlprefix\url{http://link.aps.org/doi/10.1103/PhysRevC.79.064324}.

\bibitem[{\citenamefont{Roth et~al.}(2011)\citenamefont{Roth, Langhammer,
  Calci, Binder, and Navr\'atil}}]{Roth11}
\bibinfo{author}{\bibfnamefont{R.}~\bibnamefont{Roth}},
  \bibinfo{author}{\bibfnamefont{J.}~\bibnamefont{Langhammer}},
  \bibinfo{author}{\bibfnamefont{A.}~\bibnamefont{Calci}},
  \bibinfo{author}{\bibfnamefont{S.}~\bibnamefont{Binder}}, \bibnamefont{and}
  \bibinfo{author}{\bibfnamefont{P.}~\bibnamefont{Navr\'atil}},
  \bibinfo{journal}{Phys. Rev. Lett.} \textbf{\bibinfo{volume}{107}},
  \bibinfo{pages}{072501} (\bibinfo{year}{2011}),
  \urlprefix\url{http://link.aps.org/doi/10.1103/PhysRevLett.107.072501}.

\bibitem[{\citenamefont{Dean et~al.}(2008)\citenamefont{Dean, Hagen,
  Hjorth-Jensen, Papenbrock, and Schwenk}}]{Dean08}
\bibinfo{author}{\bibfnamefont{D.~J.} \bibnamefont{Dean}},
  \bibinfo{author}{\bibfnamefont{G.}~\bibnamefont{Hagen}},
  \bibinfo{author}{\bibfnamefont{M.}~\bibnamefont{Hjorth-Jensen}},
  \bibinfo{author}{\bibfnamefont{T.}~\bibnamefont{Papenbrock}},
  \bibnamefont{and} \bibinfo{author}{\bibfnamefont{A.}~\bibnamefont{Schwenk}},
  \bibinfo{journal}{Phys. Rev. Lett.} \textbf{\bibinfo{volume}{101}},
  \bibinfo{pages}{119201} (\bibinfo{year}{2008}),
  \urlprefix\url{http://link.aps.org/doi/10.1103/PhysRevLett.101.119201}.

\bibitem[{\citenamefont{Roth and Navr\'atil}(2008)}]{Roth08-reply}
\bibinfo{author}{\bibfnamefont{R.}~\bibnamefont{Roth}} \bibnamefont{and}
  \bibinfo{author}{\bibfnamefont{P.}~\bibnamefont{Navr\'atil}},
  \bibinfo{journal}{Phys. Rev. Lett.} \textbf{\bibinfo{volume}{101}},
  \bibinfo{pages}{119202} (\bibinfo{year}{2008}),
  \urlprefix\url{http://link.aps.org/doi/10.1103/PhysRevLett.101.119202}.

\bibitem[{\citenamefont{Wildenthal}(1984)}]{Wildenthal19845}
\bibinfo{author}{\bibfnamefont{B.}~\bibnamefont{Wildenthal}},
  \bibinfo{journal}{Progress in Particle and Nuclear Physics}
  \textbf{\bibinfo{volume}{11}}, \bibinfo{pages}{5 } (\bibinfo{year}{1984}),
  ISSN \bibinfo{issn}{0146-6410},
  \urlprefix\url{http://www.sciencedirect.com/science/article/pii/0146641084900115}.

\bibitem[{\citenamefont{\^Okubo}(1954)}]{Okubo54}
\bibinfo{author}{\bibfnamefont{S.}~\bibnamefont{\^Okubo}},
  \bibinfo{journal}{Progress of Theoretical Physics}
  \textbf{\bibinfo{volume}{12}}, \bibinfo{pages}{603} (\bibinfo{year}{1954}),
  \urlprefix\url{http://ptp.ipap.jp/link?PTP/12/603/}.

\bibitem[{\citenamefont{Suzuki and Lee}(1980)}]{Suzuki80}
\bibinfo{author}{\bibfnamefont{K.}~\bibnamefont{Suzuki}} \bibnamefont{and}
  \bibinfo{author}{\bibfnamefont{S.~Y.} \bibnamefont{Lee}},
  \bibinfo{journal}{Progress of Theoretical Physics}
  \textbf{\bibinfo{volume}{64}}, \bibinfo{pages}{2091} (\bibinfo{year}{1980}),
  \urlprefix\url{http://ptp.ipap.jp/link?PTP/64/2091/}.

\bibitem[{\citenamefont{Gloeckner and Lawson}(1974)}]{Gloeckner74}
\bibinfo{author}{\bibfnamefont{D.}~\bibnamefont{Gloeckner}} \bibnamefont{and}
  \bibinfo{author}{\bibfnamefont{R.}~\bibnamefont{Lawson}},
  \bibinfo{journal}{Physics Letters B} \textbf{\bibinfo{volume}{53}},
  \bibinfo{pages}{313 } (\bibinfo{year}{1974}), ISSN \bibinfo{issn}{0370-2693},
  \urlprefix\url{http://www.sciencedirect.com/science/article/pii/0370269374903906}.

\bibitem[{\citenamefont{Moshinsky}(1959)}]{Moshinsky59}
\bibinfo{author}{\bibfnamefont{M.}~\bibnamefont{Moshinsky}},
  \bibinfo{journal}{Nuclear Physics} \textbf{\bibinfo{volume}{13}},
  \bibinfo{pages}{104 } (\bibinfo{year}{1959}), ISSN \bibinfo{issn}{0029-5582},
  \urlprefix\url{http://www.sciencedirect.com/science/article/pii/0029558259901439}.

\bibitem[{\citenamefont{Roth et~al.}(2009)\citenamefont{Roth, Gour, and
  Piecuch}}]{Roth09b}
\bibinfo{author}{\bibfnamefont{R.}~\bibnamefont{Roth}},
  \bibinfo{author}{\bibfnamefont{J.~R.} \bibnamefont{Gour}}, \bibnamefont{and}
  \bibinfo{author}{\bibfnamefont{P.}~\bibnamefont{Piecuch}},
  \bibinfo{journal}{Physics Letters B} \textbf{\bibinfo{volume}{679}},
  \bibinfo{pages}{334 } (\bibinfo{year}{2009}), ISSN \bibinfo{issn}{0370-2693},
  \urlprefix\url{http://www.sciencedirect.com/science/article/pii/S0370269309009265}.

\bibitem[{\citenamefont{Sherrill and {Schaefer III}}(1999)}]{Sherrill99}
\bibinfo{author}{\bibfnamefont{C.~D.} \bibnamefont{Sherrill}} \bibnamefont{and}
  \bibinfo{author}{\bibfnamefont{H.~F.} \bibnamefont{{Schaefer III}}}
  (\bibinfo{publisher}{Academic Press}, \bibinfo{year}{1999}),
  vol.~\bibinfo{volume}{34} of \emph{\bibinfo{series}{Advances in Quantum
  Chemistry}}, pp. \bibinfo{pages}{143 -- 269},
  \urlprefix\url{http://www.sciencedirect.com/science/article/pii/S0065327608605328}.

\bibitem[{\citenamefont{Kruse}(2012)}]{Kruse-thesis}
\bibinfo{author}{\bibfnamefont{M.}~\bibnamefont{Kruse}}, Ph.D. thesis,
  \bibinfo{school}{University of Arizona} (\bibinfo{year}{2012}),
  \bibinfo{note}{ph.D.Thesis (Advisor: B.R. Barrett)}.

\bibitem[{\citenamefont{Surj\'{a}n et~al.}(2004)\citenamefont{Surj\'{a}n,
  Rolik, Szabados, and K\"{o}halmi}}]{Surjan04}
\bibinfo{author}{\bibfnamefont{P.~R.} \bibnamefont{Surj\'{a}n}},
  \bibinfo{author}{\bibfnamefont{Z.}~\bibnamefont{Rolik}},
  \bibinfo{author}{\bibfnamefont{A.}~\bibnamefont{Szabados}}, \bibnamefont{and}
  \bibinfo{author}{\bibfnamefont{D.}~\bibnamefont{K\"{o}halmi}},
  \bibinfo{journal}{Annalen der Physik} \textbf{\bibinfo{volume}{13}},
  \bibinfo{pages}{223} (\bibinfo{year}{2004}), ISSN \bibinfo{issn}{1521-3889},
  \urlprefix\url{http://dx.doi.org/10.1002/andp.200310074}.

\bibitem[{\citenamefont{Rolik et~al.}(2003)\citenamefont{Rolik, \'{A}gnes
  Szabados, and Surj\'{a}n}}]{Rolik03}
\bibinfo{author}{\bibfnamefont{Z.}~\bibnamefont{Rolik}},
  \bibinfo{author}{\bibnamefont{\'{A}gnes Szabados}}, \bibnamefont{and}
  \bibinfo{author}{\bibfnamefont{P.~R.} \bibnamefont{Surj\'{a}n}},
  \bibinfo{journal}{The Journal of Chemical Physics}
  \textbf{\bibinfo{volume}{119}}, \bibinfo{pages}{1922} (\bibinfo{year}{2003}),
  \urlprefix\url{http://link.aip.org/link/?JCP/119/1922/1}.

\bibitem[{\citenamefont{Navr\'atil}(1995-)}]{ncsd}
\bibinfo{author}{\bibfnamefont{P.}~\bibnamefont{Navr\'atil}},
  \emph{\bibinfo{title}{No-core shell model slater determinant code (ncsd)}},
  \bibinfo{howpublished}{unpublished} (\bibinfo{year}{1995-}).

\bibitem[{\citenamefont{Vary}(1992)}]{mfd92}
\bibinfo{author}{\bibfnamefont{J.}~\bibnamefont{Vary}},
  \emph{\bibinfo{title}{The many-fermion-dynamics shell-model code}},
  \bibinfo{howpublished}{(Iowa State University),unpublished}
  (\bibinfo{year}{1992}).

\bibitem[{\citenamefont{Vary and Zheng}(1994)}]{mfd94}
\bibinfo{author}{\bibfnamefont{J.}~\bibnamefont{Vary}} \bibnamefont{and}
  \bibinfo{author}{\bibfnamefont{D.}~\bibnamefont{Zheng}},
  \emph{\bibinfo{title}{The many-fermion-dynamics shell-model code}},
  \bibinfo{howpublished}{unpublished} (\bibinfo{year}{1994}).

\bibitem[{\citenamefont{Sternberg et~al.}(2008)\citenamefont{Sternberg, Ng,
  Yang, Maris, Vary, Sosonkina, and Le}}]{mfdna}
\bibinfo{author}{\bibfnamefont{P.}~\bibnamefont{Sternberg}},
  \bibinfo{author}{\bibfnamefont{E.~G.} \bibnamefont{Ng}},
  \bibinfo{author}{\bibfnamefont{C.}~\bibnamefont{Yang}},
  \bibinfo{author}{\bibfnamefont{P.}~\bibnamefont{Maris}},
  \bibinfo{author}{\bibfnamefont{J.~P.} \bibnamefont{Vary}},
  \bibinfo{author}{\bibfnamefont{M.}~\bibnamefont{Sosonkina}},
  \bibnamefont{and} \bibinfo{author}{\bibfnamefont{H.~V.} \bibnamefont{Le}}, in
  \emph{\bibinfo{booktitle}{Proceedings of the 2008 ACM/IEEE conference on
  Supercomputing}} (\bibinfo{publisher}{IEEE Press},
  \bibinfo{address}{Piscataway, NJ, USA}, \bibinfo{year}{2008}), SC '08, pp.
  \bibinfo{pages}{15:1--15:12}, ISBN \bibinfo{isbn}{978-1-4244-2835-9},
  \urlprefix\url{http://dl.acm.org/citation.cfm?id=1413370.1413386}.

\bibitem[{\citenamefont{Maris et~al.}(2010{\natexlab{b}})\citenamefont{Maris,
  Sosonkina, Vary, Ng, and Yang}}]{Maris10}
\bibinfo{author}{\bibfnamefont{P.}~\bibnamefont{Maris}},
  \bibinfo{author}{\bibfnamefont{M.}~\bibnamefont{Sosonkina}},
  \bibinfo{author}{\bibfnamefont{J.~P.} \bibnamefont{Vary}},
  \bibinfo{author}{\bibfnamefont{E.}~\bibnamefont{Ng}}, \bibnamefont{and}
  \bibinfo{author}{\bibfnamefont{C.}~\bibnamefont{Yang}},
  \bibinfo{journal}{Procedia Computer Science} \textbf{\bibinfo{volume}{1}},
  \bibinfo{pages}{97 } (\bibinfo{year}{2010}{\natexlab{b}}), ISSN
  \bibinfo{issn}{1877-0509}, \bibinfo{note}{iCCS 2010},
  \urlprefix\url{http://www.sciencedirect.com/science/article/pii/S187705091000013X}.

\bibitem[{\citenamefont{Caurier and Nowacki}(1999)}]{antoine}
\bibinfo{author}{\bibfnamefont{E.}~\bibnamefont{Caurier}} \bibnamefont{and}
  \bibinfo{author}{\bibfnamefont{F.}~\bibnamefont{Nowacki}},
  \bibinfo{journal}{Acta Phys. Pol. B} \textbf{\bibinfo{volume}{30}},
  \bibinfo{pages}{705} (\bibinfo{year}{1999}),
  \urlprefix\url{http://th-www.if.uj.edu.pl/acta/vol30/abs/v30p0705.htm}.

\bibitem[{\citenamefont{Navr\'atil and Caurier}(2004)}]{Nav04}
\bibinfo{author}{\bibfnamefont{P.}~\bibnamefont{Navr\'atil}} \bibnamefont{and}
  \bibinfo{author}{\bibfnamefont{E.}~\bibnamefont{Caurier}},
  \bibinfo{journal}{Phys. Rev. C} \textbf{\bibinfo{volume}{69}},
  \bibinfo{pages}{014311} (\bibinfo{year}{2004}),
  \urlprefix\url{http://link.aps.org/doi/10.1103/PhysRevC.69.014311}.

\bibitem[{\citenamefont{Coon et~al.}(2012)\citenamefont{Coon, Avetian, Kruse,
  van Kolck, Maris, and Vary}}]{Coon12}
\bibinfo{author}{\bibfnamefont{S.~A.} \bibnamefont{Coon}},
  \bibinfo{author}{\bibfnamefont{M.~I.} \bibnamefont{Avetian}},
  \bibinfo{author}{\bibfnamefont{M.~K.~G.} \bibnamefont{Kruse}},
  \bibinfo{author}{\bibfnamefont{U.}~\bibnamefont{van Kolck}},
  \bibinfo{author}{\bibfnamefont{P.}~\bibnamefont{Maris}}, \bibnamefont{and}
  \bibinfo{author}{\bibfnamefont{J.~P.} \bibnamefont{Vary}},
  \bibinfo{journal}{Phys. Rev. C} \textbf{\bibinfo{volume}{86}},
  \bibinfo{pages}{054002} (\bibinfo{year}{2012}),
  \urlprefix\url{http://link.aps.org/doi/10.1103/PhysRevC.86.054002}.

\bibitem[{\citenamefont{Furnstahl et~al.}(2012)\citenamefont{Furnstahl, Hagen,
  and Papenbrock}}]{Furnstahl12}
\bibinfo{author}{\bibfnamefont{R.~J.} \bibnamefont{Furnstahl}},
  \bibinfo{author}{\bibfnamefont{G.}~\bibnamefont{Hagen}}, \bibnamefont{and}
  \bibinfo{author}{\bibfnamefont{T.}~\bibnamefont{Papenbrock}},
  \bibinfo{journal}{Phys. Rev. C} \textbf{\bibinfo{volume}{86}},
  \bibinfo{pages}{031301} (\bibinfo{year}{2012}),
  \urlprefix\url{http://link.aps.org/doi/10.1103/PhysRevC.86.031301}.

\bibitem[{\citenamefont{Navr\'atil and Quaglioni}(2011)}]{Nav11-da}
\bibinfo{author}{\bibfnamefont{P.}~\bibnamefont{Navr\'atil}} \bibnamefont{and}
  \bibinfo{author}{\bibfnamefont{S.}~\bibnamefont{Quaglioni}},
  \bibinfo{journal}{Phys. Rev. C} \textbf{\bibinfo{volume}{83}},
  \bibinfo{pages}{044609} (\bibinfo{year}{2011}),
  \urlprefix\url{http://link.aps.org/doi/10.1103/PhysRevC.83.044609}.

\bibitem[{\citenamefont{Navr\'atil et~al.}(2010)\citenamefont{Navr\'atil, Roth,
  and Quaglioni}}]{PhysRevC.82.034609}
\bibinfo{author}{\bibfnamefont{P.}~\bibnamefont{Navr\'atil}},
  \bibinfo{author}{\bibfnamefont{R.}~\bibnamefont{Roth}}, \bibnamefont{and}
  \bibinfo{author}{\bibfnamefont{S.}~\bibnamefont{Quaglioni}},
  \bibinfo{journal}{Phys. Rev. C} \textbf{\bibinfo{volume}{82}},
  \bibinfo{pages}{034609} (\bibinfo{year}{2010}),
  \urlprefix\url{http://link.aps.org/doi/10.1103/PhysRevC.82.034609}.

\end{thebibliography}

\end{document}